\documentclass{article}

\usepackage{arxiv}

\usepackage[utf8]{inputenc}            
\usepackage[T1]{fontenc}               
\usepackage[hidelinks]{hyperref}       
\usepackage{url}                       
\usepackage{booktabs}                  
\usepackage{amsfonts}                  
\usepackage{nicefrac}                  
\usepackage{microtype}                 
\usepackage{lipsum}		               
\usepackage{tabularx}
\usepackage{rotating}
\usepackage{amsmath} 
\usepackage{siunitx}
\usepackage{multirow}
\usepackage{makecell}
\usepackage{siunitx} 
\newcolumntype{L}{>{\raggedright\arraybackslash}X}

\usepackage{graphicx}
\usepackage[numbers,sort&compress]{natbib}
\usepackage{doi}
\usepackage{caption}

\title{The Effect of Design Thinking on Creative \& Innovation Processes: An Empirical Study Across Different Design Experience Levels}

\author{ 
    \includegraphics[scale=0.06]{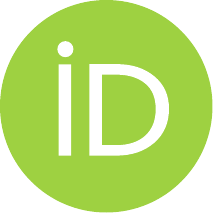}\hspace{1mm}Yuxin Zhang
    \thanks{Yuxin Zhang, PhD in Design, focuses on industrial design, material experience, affective engineering, and ergonomics. For further academic details, please refer to his ORCID profile: https://orcid.org/0000-0001-8100-3649} \\
    Academy of Arts \& Design \\
    Tsinghua University \\
    Haidian  District, Beijing, 100084, China \\
    \texttt{zhangyuxin@mail.tsinghua.edu.cn} \\
    \And
	Fan Zhang \\
	Academy of Arts \& Design \\
	Tsinghua University \\
	Haidian  District, Beijing, 100084, China \\
	\texttt{zhang-f24@mails.tsinghua.edu.cn} \\
}

\renewcommand{\shorttitle}{The Effect of Design Thinking on Creative \& Innovation Processes}

\hypersetup{
pdftitle={A template for the arxiv style},
pdfsubject={q-bio.NC, q-bio.QM},
pdfauthor={David S.~Hippocampus, Elias D.~Striatum},
pdfkeywords={First keyword, Second keyword, More},
}

\begin{document}
\maketitle

\begin{abstract}
This study employs linear regression and structural equation modeling to explore how Thinking Skills, Design Thinking, Creative Self-Efficacy (CSE), and Collective Creative Efficacy (CCE) drive Design Creativity \& Innovation, and analyzes the structural stability of the model across different levels of experience. Path analysis results indicate that the four Design Thinking Skills, Problem-driven Design ($\beta = 0.198$, $p < 0.01$), Information-driven Design ($\beta = 0.241$, $p < 0.001$), Solution-driven Design ($\beta = 0.227$, $p < 0.001$), and Knowledge-driven Design ($\beta = 0.263$, $p < 0.001$) all significantly and positively influence Design Thinking. Furthermore, Design Thinking has a significant positive predictive effect on Design Creativity \& Innovation ($\beta = 0.286$, $p < 0.001$). Mediation analysis confirms three significant mediation paths: the CSE mediation path ($\beta = 0.128$, $p < 0.001$), the CCE mediation path ($\beta = 0.073$, $p < 0.01$), and the "$\text{CSE} \rightarrow \text{CCE}$" chain mediation path ($\beta = 0.025$, $p < 0.01$). Multi-group comparison results reveal significant differences between the student and professional groups under the full equivalence model. After relaxing specific constraints, there were no significant differences between the nested models of the baseline model, partial measurement invariance, structural weight invariance, and structural covariance invariance. These findings elucidate the multi-dimensional pathways of Design Creativity \& Innovation, providing a robust empirical basis for optimizing differentiated pedagogical models and professional practice guidelines.
\end{abstract}


\section{Introduction}
In the field of design, creativity and innovation serve as key evaluative benchmarks and the core impetus for advancing design progress while addressing complex challenges \citep{amabileCreativityContextUpdate2018}. These constructs encompass not only the generation of novel ideation but also the reframing of conventional mental frameworks to explore unprecedented solutions \citep{dorstCoreDesignThinking2011}, as well as their subsequent actualization to facilitate technological breakthroughs \citep{bareghehMultidisciplinaryDefinitionInnovation2009}. Within design practice, the realization of creativity and innovation necessitates a sophisticated dialectic between analytical rigor and heuristic synthesis, predicated upon the strategic configuration of thinking skills to reconcile multifaceted systemic constraints \citep{oxmanEducatingDesignerlyThinker1999}. Against this backdrop, Design Thinking has emerged as a critical methodological paradigm that synthesizes these diverse thinking skills through a human-centered and iterative logic, thereby providing a structured framework for the co-evolution of problem and solution spaces to catalyze Design Creativity \& Innovation \citep{buchananWickedProblemsDesign1992}. However, the efficacy of this methodology is not universally uniform; rather, it is profoundly contingent upon the designer’s internal cognitive agency. Consequently, establishing a structural model to elucidate the mechanisms between Thinking Skills, Design Thinking, and Design Creativity \& Innovation is crucial for elucidating the underlying paths through which individual cognitive capacities are restructured under methodological guidance to drive high-level innovative performance.

This cognitive transition from Thinking Skills to Design Creativity \& Innovation does not occur in a vacuum but is increasingly situated within complex interdisciplinary practices. As the design paradigm shifts from individualistic pursuits toward collective intelligence, the integration of collaboration and shared wisdom has become indispensable \citep{kleinsmannBarriersEnablersCreating2008}. Within this collaborative framework, the actualization of a designer’s cognitive potential is closely linked to their self-belief systems, specifically Creative Self-Efficacy and Collective Creative Efficacy. While Creative Self-Efficacy represents an individual’s internal conviction in their ability to produce creative outcomes \citep{tierneyCreativeSelfEfficacyIts2002}, Collective Creative Efficacy functions as the shared belief in a team’s joint capability to execute innovative actions \citep{chengAntecedentsCollectiveCreative2014}. Although understanding how Creative Self-Efficacy and Collective Creative Efficacy interact with Design Thinking to promote Design Creativity \& Innovation is essential for advancing design theory, it is still unclear how these two forms of efficacy mediate the process through which methodology is applied to achieve innovative outcomes. Existing research has yet to reach consensus on the structural role between these two factors in complex design environments, and further empirical research is urgently needed to clarify this relationship \citep{mathisenCreativeSelfefficacyIntervention2009}.

The complexity of these path mechanisms is further compounded by the evolution of design cognition from academic training to professional practice. While educational environments typically prioritize the expression of self-expression and the stimulation of divergent inspiration, professional socialization necessitates the internalization of these processes into systematic and domain-specific heuristic methods \citep{christiaansCreativityDesignEngineering2005}. This fundamental shift raises significant questions regarding the structural stability of the innovation mechanism as a designer progresses from novice to expert. Although research on design expertise has highlighted distinct behavioral patterns between different developmental stages, it remains unclear whether the interplay between Thinking Skills, Design Thinking, Creative Efficacy, and Design Creativity \& Innovation maintains a consistent predictive structure across these populations \citep{atmanEngineeringDesignProcesses2007, tierneyCreativeSelfefficacyDevelopment2011}. Consequently, exploring the cognitive variations between students and seasoned professionals is essential to determine whether the proposed innovation path represents a universal mechanism or whether it undergoes a profound reconfiguration during the development of professional expertise.

This study aims to explore the path mechanisms between Thinking Skills, Design Thinking, Creative Efficacy, and Design Creativity \& Innovation through linear regression and structural equation modeling analysis. It also seeks to evaluate the mediating role of creative efficacy and assess the structural stability of the innovation mechanism across different developmental stages through multi-group comparison. Ultimately, by elucidating the stability and variation of these structural paths, this study aims to provide a robust theoretical foundation for optimizing the development of expertise from education to professional practice.

\section{Foundational Components in the Theoretical Framework}
\subsection{Thinking Skills in Design}
Based on established design research, we find that design cognition is fundamentally structured around a comprehensive expertise model involving eight core activities: (1) gathering data, (2) assessing data validity, (3) identifying constraints and requirements, (4) modeling behaviors and environments, (5) defining problems and possibilities, (6) generating partial solutions, (7) evaluating solutions, and (8) assembling coherent outcomes \citep{krugerSolutionDrivenProblem2006}. In practice, these activities integrate into four primary cognitive strategies: Problem-driven, Information-driven, Solution-driven, and Knowledge-driven Design \citep{krugerSolutionDrivenProblem2006}. In the context of this study, these cognitive strategies are viewed as a versatile set of Thinking Skills that designers deploy to navigate complex creative challenges. This skill-based perspective encompasses analytical skills for problem deconstruction, evidence-based skills for information synthesis, generative skills for solution-led reframing, and associative skills for knowledge transfer. While every designer possesses this full cognitive repertoire, the relative intensity and configuration of these thinking skills vary individually, creating a unique cognitive profile for each person.

\subsubsection{Problem-driven Design (PD)}
Problem-driven Design adopts the principle of "definition before adaptation," focusing on a rigorous logical deconstruction of the task. As observed by Kruger and Cross \citep{krugerSolutionDrivenProblem2006}, these designers prioritize clarifying goal boundaries and constraints through iterative interpretations, effectively preventing premature leaps into solution generation. Their information strategy follows a "minimalist" approach, selecting only data directly relevant to problem definition to avoid cognitive overload. This systematic path yields a highly specified problem structure, leading to focused creative outputs. Consequently, the evaluation remains strictly "requirement-orientated," ensuring that the outcomes align perfectly with pre-established constraints \citep{luRelationshipStudentDesign2015}.

\subsubsection{Information-driven Design (ID)}
Information-driven Design treats "evidence-based construction" as the key driver, where the systematic acquisition of external data drives the process. Kruger and Cross \citep{krugerSolutionDrivenProblem2006} found that this approach creates a meticulously detailed problem space, viewing the brief as a blueprint for identifying external "pointers." Lu \citep{luRelationshipStudentDesign2015} further notes that this orientation is prevalent among lower-grade students, reflecting a reliance on process knowledge such as data structuring when domain experience is limited. Within this paradigm, solutions are direct projections of gathered evidence, resulting in fewer but more factually robust options. The assessment prioritizes alignment with market trends and technical standards, embodying a fusion of rationality and evidence.

\subsubsection{Solution-driven Design (SD)}
Solution-driven Design operates on "generation first, followed by problem reframing," fundamentally breaking away from traditional linear models. Designers intentionally keep the problem "ill-defined" to expand the creative search space and bypass exhaustive preliminary analysis \citep{krugerSolutionDrivenProblem2006}. This strategy relies on the rapid retrieval of structured knowledge and intuition from memory, enabling an earlier entry into the synthesis phase. Information gathering remains highly selective, occurring only when specific concepts require empirical validation. Consequently, evaluation transforms into a dynamic tool for redefining problem boundaries, where emerging ideas reconfigure initial constraints through an open-ended exploration of possibilities \citep{krugerSolutionDrivenProblem2006, luRelationshipStudentDesign2015}.

\subsubsection{Knowledge-driven Design (KD)}
Knowledge-driven Design centers on "experience adaptation," where designers compare current assignments with similar cases stored in their memory. Their unique information strategy focuses on "filling knowledge gaps," gathering new data only when existing expertise proves insufficient, thereby shortening the exploration path \citep{krugerSolutionDrivenProblem2006}. By linking task requirements to existing knowledge frameworks, this approach narrows the problem space rapidly, ensuring both reliability and efficiency. Characterized by a "long analysis, short synthesis" cycle, Knowledge-driven Design avoids blind exploration by providing a solid foundation of prior experience, balancing efficiency with credible design outcomes \citep{luRelationshipStudentDesign2015}.

\subsection{Design Thinking (DT)}
Design Thinking is conceptualized as a systematic, human-centered innovation paradigm that integrates interdisciplinary knowledge to navigate complex problem spaces \citep{carlgrenDesignThinkingExploring2014}. Scholarly evidence suggests that the effective deployment of this paradigm is fundamentally contingent upon cognitive skills. Specifically, the impetus of empathetic thinking and the cognitive flexibility to oscillate between divergent and convergent thinking determine the depth and breadth of exploring the solution space \citep{dymEngineeringDesignThinking2005}. By embedding thinking skills into the innovation trajectory, Design Thinking establishes a structured learning process that prioritizes user experience through continuous feedback loops \citep{beckmanInnovationLearningProcess2007}, while driving iterative refinement through rigorous prototyping and experimentation \citep{elsbachDesignThinkingOrganizational2018}. Such methodologies serve as a vital engine for enhancing Design Creativity \& Innovation by strengthening the organizational capacity to identify unarticulated needs and convert them into competitive advantages. This direct link is manifested in how design thinking optimizes the efficiency of problem-solution matching, thereby significantly elevating the success rate and output efficacy of design-driven innovation \citep{liedtkaPerspectiveLinkingDesign2015}.

\subsection{Creative Efficacy at Both Individual and Collective Levels}
\subsubsection{Creative Self-Efficacy (CSE)}
In the field of design, Creative Self-Efficacy refers to a designer's subjective belief in their ability to generate novel, functional, and aesthetically coherent solutions to complex problems, while also serving as a specialized psychological motivation that empowers designers to navigate the inherent ambiguity of the creative process \citep{tierneyCreativeSelfEfficacyIts2002, lambMeasuringCreativeSelfefficacy2025}. Unlike general professional confidence, design-specific Creative Self-Efficacy emphasizes the cognitive agility required for problem deconstruction and iterative reframing. It is significantly influenced by an individual’s innovative cognitive style and their capacity for self-regulation, including activities such as planning and monitoring task progress \citep{beeftinkBeingSuccessfulCreative2012}. High levels of Creative Self-Efficacy encourage designers to persevere through setbacks and maintain creative momentum when faced with conflicting constraints. Furthermore, research indicates that Creative Self-Efficacy acts as a critical mediator between a designer’s cognitive orientation and their professional success \citep{beeftinkBeingSuccessfulCreative2012}. By fostering a robust sense of creative agency, Creative Self-Efficacy enables individuals to translate abstract design skills into tangible, impactful outcomes. Consequently, cultivating Creative Self-Efficacy is essential for developing professional expertise and achieving excellence within the design industry \citep{lambMeasuringCreativeSelfefficacy2025}.

\subsubsection{Collective Creative Efficacy (CCE)}
Collective Creative Efficacy, also referred to as team creative efficacy, represents the shared conviction among design team members regarding their joint capacity to produce novel and functional outcomes \citep{shinWhenEducationalSpecialization2007}. Within the field of design, Collective Creative Efficacy is viewed as a psychological state resulting from the dynamic interplay between individual motivation and team-level interactions during complex tasks \citep{chenSystemsTheoryMotivated2006, jablokowInvestigatingInfluenceDesigners2019}. It is a social-behavioral attribute that enables teams to coordinate cognitive resources and integrate diverse expertise, working together towards a unified creative goal \citep{chenSystemsTheoryMotivated2006, shinWhenEducationalSpecialization2007}. In collaborative design environments, high levels of Collective Creative Efficacy enable designers to leverage educational heterogeneity to achieve higher levels of innovation \citep{shinWhenEducationalSpecialization2007, jablokowInvestigatingInfluenceDesigners2019}. Ultimately, Collective Creative Efficacy sustains team persistence and cognitive agility, ensuring that the team's potential is effectively transformed into tangible and aesthetically coherent solutions.

\subsection{Design Creativity \& Innovation (DCI)}
Design Creativity \& Innovation is regarded as a multidimensional capability that integrates aesthetic expression with functional problem-solving \citep{amabileCreativityContextUpdate2018}. Rooted in the systematic transformation of "existing situations into preferred ones," it operates within the rigorous constraints of technical feasibility and market viability \citep{simonSciencesArtificial2019}. Consequently, Design Creativity \& Innovation serves as the determinant factor for design success, acting as the strategic nexus that translates creative potential into tangible market value. The integration of Design Creativity \& Innovation ensures that design transcends peripheral aesthetic exercise, becoming a systemic driver capable of converting abstract insights into high-impact innovations with predictable excellence \citep{martinDesignBusinessWhy2009}. Crucially, Design Creativity \& Innovation facilitates the alignment of technical feasibility with user desirability, ensuring that creative outputs are not merely "novel" but "strategically valuable" within complex systemic constraints. By leveraging design cognitive skills as a strategic asset, Design Creativity \& Innovation mitigates inherent risks in the innovation cycle and significantly elevates the success rate of design-driven solutions. Ultimately, it is the presence of Design Creativity \& Innovation that secures the utility of outputs, ensuring that vision is effectively translated into a cumulative record of professional achievement and competitive advantage \citep{carsonReliabilityValidityFactor2005}.

\section{Theoretical Model of Design Thinking Driven Creativity \& Innovation}
\subsection{Conceptual Framework}
\begin{figure}[htbp]
	\centering
		\includegraphics[scale=0.2]{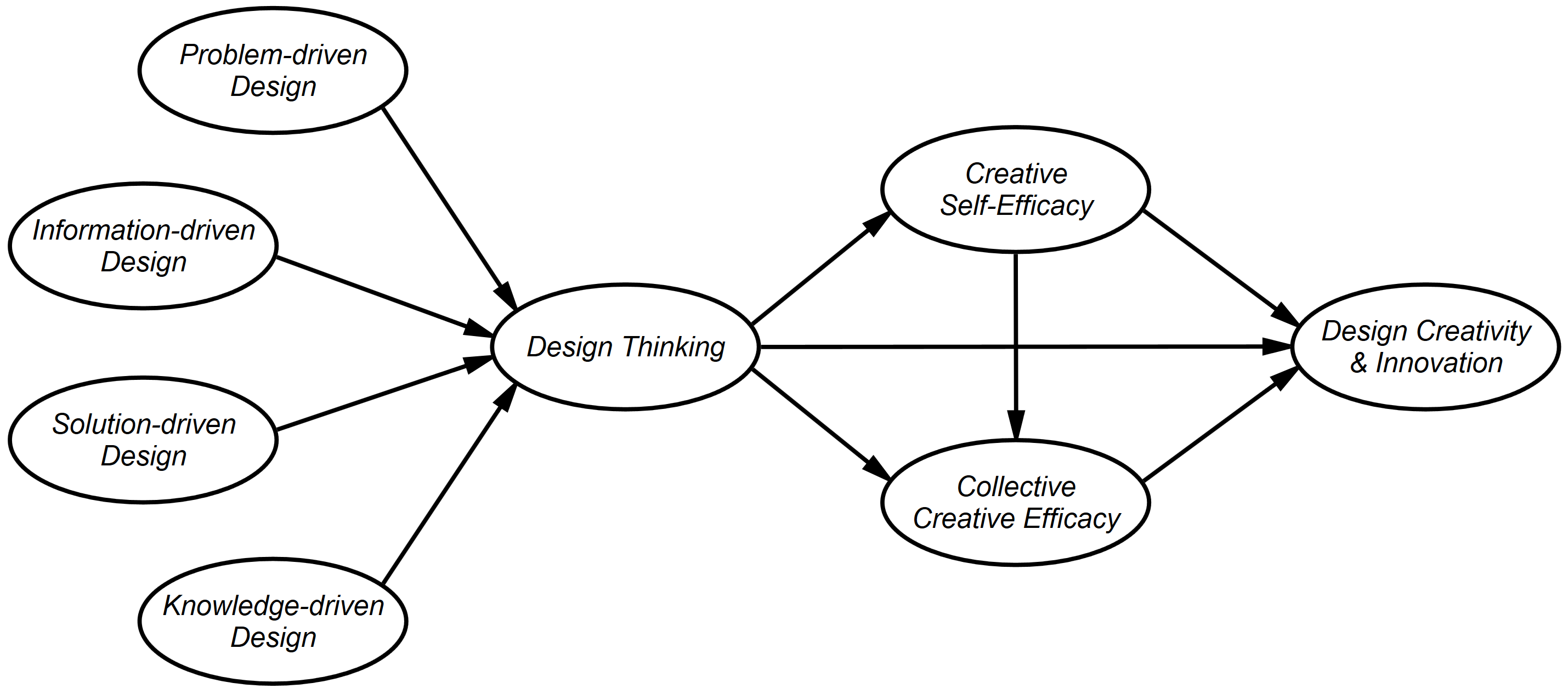}
	\caption{Design Thinking Driven Creativity and Innovation Model}
	\label{fig: Theoretical Model}
\end{figure}

As showed in Figure \ref{fig: Theoretical Model}, this study proposes an integrated theoretical framework that bridges cognitive orientations, methodological processes, psychological agency and the advancement of creativity and innovation. The model posits that the four cognitive strategies—Problem-driven, Information-driven, Solution-driven, and Knowledge-driven Design—function as the foundational cognitive repertoire necessary for navigating complex design tasks \citep{krugerSolutionDrivenProblem2006}. Design Thinking is defined as a structured, iterative learning process that bridges abstract reasoning with concrete practice. By facilitating the continuous refinement of design concepts through feedback loops, it provides a systematic pathway for driving Design Creativity \& Innovation \citep{beckmanInnovationLearningProcess2007}. 

Furthermore, the framework highlights the role of psychological agency in regulation and reshaping. It proposes three evolutionary paths: (1) individual confidence, (2) collective collaboration confidence, and (3) the progressive evolutionary path from individual confidence to collective collaboration confidence. At the individual level, Design Thinking provides the necessary structured support for complex creative tasks, influencing the designer’s Creative Self-Efficacy \citep{tierneyCreativeSelfEfficacyIts2002}. Meanwhile, Design Thinking integrates and synchronizes the diverse professional knowledge of cross-disciplinary teams toward a unified innovation goal, thereby facilitating the formation of Collective Creative Efficacy \citep{chenSystemsTheoryMotivated2006}. Crucially, the model establishes a unique driving path by aggregating individual confidence into cohesive collective power, effectively bridging the gap between individual potential and team-level performance \citep{chenSystemsTheoryMotivated2006}. These efficacy states serve as critical motivational mediators, ensuring that raw cognitive potential is effectively mobilized and directed toward goal-oriented innovation outputs \citep{gongEmployeeLearningOrientation2009}. Ultimately, the synergy between these cognitive orientations, thinking paradigms, and psychological drivers consolidates the systemic integrity of the innovation model at multiple levels, ensuring that output outcomes are both novel and functionally effective \citep{amabileDynamicComponentialModel2016}.

\subsection{Research Logic and Path Delineation}
To empirically validate the proposed framework, this study employs a four-stage analytical strategy. First, the study evaluates the predictive influence of a designer’s cognitive repertoire, which includes four core strategies: Problem-driven, Information-driven, Solution-driven, and Knowledge-driven Design. It explores how these Thinking Skills affect Design Thinking. Second, we examine the role of Design Thinking in promoting Design Creativity \& Innovation. Third, the study uncovers the psychological mechanisms within the model by testing how Design Thinking drives Design Creativity \& Innovation through the activation of Creative Self-Efficacy and Collective Creative Efficacy as mediating factors. Finally, we ensure the structural stability of the model through multi-group invariance testing. By comparing student and professional groups, this analysis determines whether the model’s mechanisms remain consistent across different levels of expertise.

\section{Method and Data Process}
\subsection{Method}
This study adopts a mixed methods approach, combining quantitative research methods such as regression analysis and structural equation modeling with qualitative research through semi-structured interviews, to explore the research problem from multiple perspectives. In the exploratory data analysis phase, composite variables were generated by calculating the arithmetic mean of each construct, followed by preliminary regression analysis to verify the hypothesized relationships in the theoretical model. Next, confirmatory factor analysis (CFA) was used to assess the composite reliability (CR) and average variance extracted (AVE) of the measurement model. After ensuring the measurement quality met the required standards, a structural equation model (SEM) was constructed. To further explore the mediating effect of psychological agency, the study performed mediation effect analysis using a bootstrap procedure with 5,000 resamples. The significance of the Creative Self-Efficacy, Collective Creative Efficacy, and their chain mediation effect was evaluated by calculating percentile confidence intervals (Percentile CI) and bias-corrected confidence intervals (Bias-corrected CI). Following this, a multi-group comparison analysis was conducted for the student and professional groups. Invariance testing was performed through hierarchical tests of form, measurement item loadings, structural weights, and covariance to identify significant differences in path coefficients across design experience. Finally, the quantitative analysis results were integrated with qualitative insights from semi-structured interviews, followed by an in-depth discussion of the research findings.

We conducted exploratory data analysis using R Project 4.5.2 and constructed the Structural Equation Model, while performing mediation effect analysis and multi-group comparison analysis, utilizing libraries lavaan, semPlot, and lavaanPlot. Additionally, we employed Python 3.14.0, incorporating libraries numpy, pandas, matplotlib, seaborn, and sklearn for data visualization.

\subsection{Measurement Items}
The measurement items used in this study was developed by adapting established scales and synthesizing existing theoretical frameworks to ensure both content validity and contextual relevance. All constructs were operationalized using a 7-point Likert scale, ranging from 1 (strongly disagree) to 7 (strongly agree). To ensure a parsimonious items and minimize participant cognitive burden, a panel of experts was invited to evaluate the relevance and clarity of the initial items. Specifically, consistent with the scale development principles suggested by Hinkin \citep{hinkinReviewScaleDevelopment1995}, five highly representative items were selected for each of the eight constructs based on expert consensus and significance in theoretical model (see Table \ref{tab: Questionnaire}). These measurement items were operationalized across four primary theoretical domains:

\begin{itemize}
    \item Design Cognition Strategies: Drawing upon the seminal taxonomy of Kruger and Cross \citep{krugerSolutionDrivenProblem2006}, design cognition was categorized into four distinct constructs: Problem-driven, Information-driven, Solution-driven, and Knowledge-driven Design. These constructs were measured using scales adapted from Lu \citep{luRelationshipStudentDesign2015}, which capture how designers prioritize their mental focus and allocate cognitive resources toward different strategic orientations during problem-solving.
    \item Design Thinking: The Design Thinking construct integrates multidimensional mental orientations with iterative behavioral processes. The measurement items were synthesized from the Design Thinking Mindset scale proposed by Vignoli et al. \citep{vignoliDesignThinkingMindset2023} and the Scale of Design Thinking for Teaching developed by Cai and Yang \citep{caiDevelopmentValidationScale2023}, capturing both the underlying design attitude and the procedural stages of design action.
    \item Creative Efficacy: This study assessed creative confidence at both individual and collective levels. Creative Self-Efficacy items were adapted from the instrument developed by Lamb et al. \citep{lambMeasuringCreativeSelfefficacy2025} and the Engineering Design Self-Efficacy scale by Carberry et al. \citep{carberryMeasuringEngineeringDesign2010}, primarily focusing on the confidence to execute specific design tasks and generate innovative concepts. Collective Creative Efficacy items were derived from the models of Cheng and Yang \citep{chengAntecedentsCollectiveCreative2014} as well as Zou et al. \citep{zouCognitiveCharacteristicsInnovation2023a} to evaluate the shared belief among team members regarding their group's creative synergy and potential for innovation output.
    \item Design Creativity \& Innovation: The Design Creativity \& Innovation construct evaluates through the two dimensions of creative self-perception and product breakthrough. These items draw upon the self-perceived creativity framework of Kreitler and Casakin \citep{kreitlerSelfPerceivedCreativityPerspective2009a} and further incorporate the product innovativeness dimensions established by McNally et al. \citep{mcnallyProductInnovativenessDimensions2010}, assessing the degree of novelty and technological/marketing discontinuity in the resulting design solutions.
\end{itemize}

\subsection{Participants and Experimental Environment}
\begin{figure}[htbp]
	\centering
		\includegraphics[scale=.248]{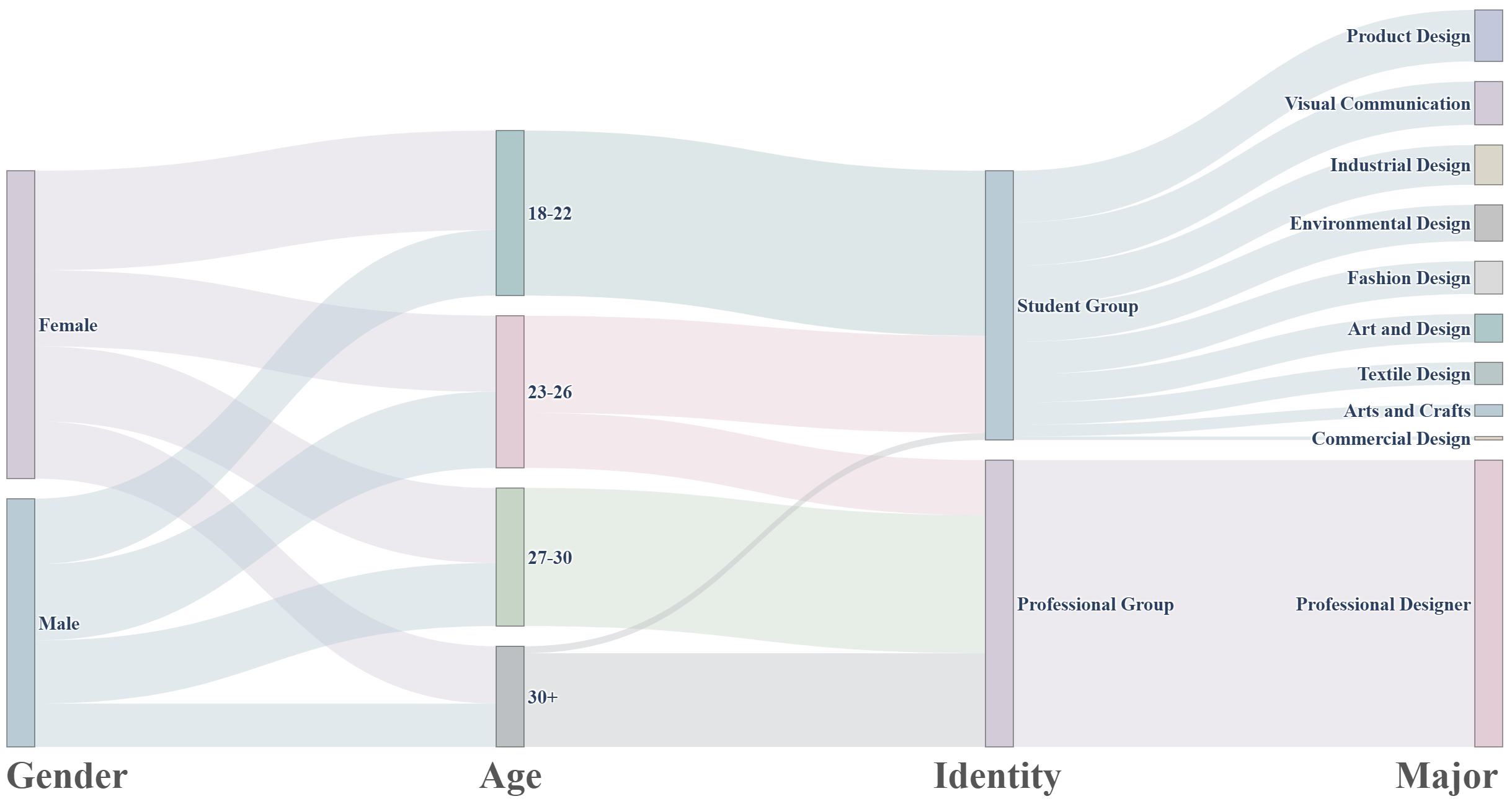}
	\caption{Distribution and Progression of Participant Demographics across Four Key Categorical Variables: Gender, Age, Identity, and Major}
	\label{fig: distribution and progression}
\end{figure}

We collected a total of 475 valid questionnaires, with data obtained through two channels: on-site completion and online surveys. Among the participants, 230 were university students, including undergraduates, master's students, and doctoral students, forming the "Student Group." The members of the Student Group mainly came from the following disciplines: product design, visual communication, industrial design, environmental design, fashion design, art and design, textile design, art and craft, and commercial design. The remaining 245 participants were professional designers who had graduated and were currently working in design-related fields, forming the "Professional Group." Due to the broad range of design fields in which the Professional Group members are involved, we did not collect detailed information regarding their specific professions. In terms of gender distribution, there were 263 male participants and 212 female participants. Regarding age distribution, 141 participants were aged between 18-22, 130 were aged between 23-26, 118 were aged between 27-30, and 86 were over 30 years old. The detailed distribution is shown in Figure \ref{fig: distribution and progression}.

During the questionnaire completion process, uniform instructions were provided to all participants to ensure consistency between on-site and online responses. The questionnaires were completed anonymously and were not linked to participants' academic performance, ensuring both the protection of their privacy and the objectivity of the data \citep{podsakoffCommonMethodBiases2003}. These procedural remedies were implemented to reduce evaluation apprehension and minimize social desirability bias. Additionally, the questionnaire design included both positively and negatively worded questions to mitigate response sets and enhance the validity and reliability of the data \citep{podsakoffSourcesMethodBias2012}. Prior to conducting the data analysis, reverse-scoring was performed on negatively worded items to ensure the accuracy and consistency of the dataset.

\subsection{Data processing and analysis}
\subsubsection{Data Exploration}
In the data exploration phase, the study employed multidimensional visualization techniques. First, a hierarchical clustering heatmap was used to globally examine all observed variables, with Pearson correlation coefficients employed to depict the linear relationships between variables. A dendrogram was also incorporated to identify the underlying aggregation logic of the variables, aiming to visually assess the relationships between each evaluation item and validate the reasonableness of the dimensional divisions. 

Subsequently, the study employed linear regression to analyze the following two sets of paths:

\begin{itemize}
  \item The linear regression relationship between Problem-driven Design, Information-driven Design, Solution-driven Design, Knowledge-driven Design, and Design Thinking;
  \item The linear regression relationship between Design Thinking and Design Creativity \& Innovation.
\end{itemize}

To achieve this, the study utilized linear regression plots with marginal distributions. By comparing the fitting curves and coefficient of determination ($R^2$) between the "Student Group" and the "Professional Group," the study explored the heterogeneity in the strength of variable relationships between the two groups. At the same time, histograms and kernel density curves along the axis edges were used to simultaneously examine the distribution shapes and central tendencies of the data for each group. This progressive exploration not only validated the data quality and distribution characteristics but also laid an empirical foundation for further statistical inference and model development.

\subsubsection{Structural Equation Model}
This study then conducted data analysis using Structural Equation Modeling. Structural Equation Modeling is a statistical tool used to explore complex relationships between variables, with its core advantage being the ability to simultaneously handle multiple causal paths and effectively model latent variables \citep{bollenStructuralEquationsLatent1989a}. In terms of parameter estimation, the Maximum Likelihood method was employed. This method estimates model parameters by maximizing the likelihood function of the observed data, aiming to find parameter values that best fit the actual data. The advantage of the Maximum Likelihood method lies in its strong statistical efficiency, as it can generally provide consistent estimates, especially when applied to complex models and large sample sizes \citep{bollenStructuralEquationsLatent1989a}.

To assess the model's goodness of fit, this study employed a multi-dimensional of indices: the Chi-square to degrees of freedom ratio ($\chi^2/df$), incremental fit indices (including GFI, AGFI, CFI, and TLI), and absolute fit indices (RMSEA and SRMR). The specific evaluation criteria follow established academic standards: a $\chi^2/df$ within the range of 1 to 3 indicates an appropriate balance between model simplicity and data fit. For the incremental fit indices, values exceeding 0.90 signify an ideal fit. For the absolute fit indices, RMSEA and SRMR measure the discrepancy between the model’s predictions and the actual data. If both indices are below 0.08, it indicates that the model's fit errors are relatively small, and the fit is good \citep{huCutoffCriteriaFit1999}. This comprehensive assessment provided rigorous validation for the model's structure and facilitated subsequent refinements.

To ensure the rigor of the study, this research conducted a reliability and validity test on the model using composite reliability and average variance extracted. According to academic conventions, composite reliability should be greater than 0.700 to ensure the internal consistency of the latent variables, and average variance extracted should be greater than 0.500 to demonstrate adequate variance explanatory power and convergent validity \citep{fornellEvaluatingStructuralEquation1981}.

\subsubsection{Mediation Analysis}
Following the recommendations of Hayes \citep{hayesIntroductionMediationModeration2018}, this study utilized a bootstrapping procedure with 5,000 resamples to evaluate the mediation effects. As a non-parametric approach superior to the traditional Sobel test, bootstrapping bypasses the strict requirement for a normally distributed indirect effect by constructing an empirical distribution directly from the dataset. To ensure the robustness of the findings, we reported 95\% confidence intervals based on both percentile and bias-corrected methods \citep{mackinnonConfidenceLimitsIndirect2004}. This dual-validation procedure significantly enhances statistical power and yields more reliable inferences \citep{preacherAsymptoticResamplingStrategies2008a}. In the analysis, we focused on three specific mediation paths:

\begin{itemize}
    \item Individual Mediation Path: Design Thinking $\rightarrow$ Creative Self-Efficacy $\rightarrow$ Design Creativity \& Innovation, reflecting the mediating role of individual beliefs;
    \item Collective Mediation Path: Design Thinking $\rightarrow$ Collective Creative Efficacy $\rightarrow$ Design Creativity \& Innovation, revealing the mediating role of team collaborative beliefs;
    \item Chain Mediation Path: Design Thinking $\rightarrow$ Creative Self-Efficacy $\rightarrow$ Collective Creative Efficacy $\rightarrow$ Design Creativity \& Innovation, aimed at capturing how individual efficacy influences the dependent variable through the enhancement of collective efficacy.
\end{itemize}

\subsubsection{Multi-Group Analysis}
In the multi-group analysis, the sample was categorized into two groups: the Student Group ($N = 230$) and the Professional Group ($N = 245$). To assess the invariance across these groups, four hierarchical nested models (M1--M4) were constructed. Regarding the decision criteria for model equivalence, this study adopted the traditional Chi-square difference test ($\Delta \chi^2$) as the primary benchmark \citep{joreskogSimultaneousFactorAnalysis1971}. This methodological choice was justified by several critical considerations: First, both groups exceeded the recommended threshold of 200 cases, which ensured stable parameter estimation and sufficient statistical power for the structural equation modeling analysis \citep{klinePrinciplesPracticeStructural2016}. Second, given that the total sample size was of a moderate scale ($N = 475$), it effectively mitigated the risk of inflated Chi-square values—a common issue in excessively large samples ($N > 1000$) that often leads to "false significance" or the over-rejection of the null hypothesis for trivial discrepancies \citep{hairMultivariateDataAnalysis2010}. Therefore, while contemporary research frequently relies on descriptive fit indices (such as $\Delta$CFI) \citep{cheungEvaluatingGoodnessofFitIndexes2002a}, the Chi-square difference test remains the most statistically rigorous method in this study \citep{bollenStructuralEquationsLatent1989a}. The four specific hierarchical nested models are defined as follows:

\begin{enumerate}
    \item Configural Invariance Model (M1): As the baseline model, M1 allows all parameters to be freely estimated across groups. It tests whether the two groups share the same basic model configuration. Acceptable fit of M1 is a prerequisite for subsequent invariance testing.
    \item Measurement Weights Invariance Model (M2): This model imposes equality constraints on all factor loadings. It verifies metric invariance, ensuring that the measurement scale is equivalent and the underlying conceptual meaning of each construct remains consistent across the two groups.
    \item Structural Weights Invariance Model (M3): This model imposes equality constraints on all structural path coefficients. It tests structural invariance, identifying whether the strength of relationships between variables differs across groups.
    \item Structural Covariances Invariance Model (M4): This model imposes equality constraints on the variances and covariances of latent variables. It evaluates the overall stability of the latent factor structure and the consistency of relationships among the constructs.
\end{enumerate}

If the chi-square difference test ($\Delta \chi^2$) in the aforementioned nested comparisons yields a significant result ($p < 0.05$), the hypothesis of full invariance is rejected. Consequently, a systematic procedure is implemented to identify the sources of non-invariance and establish a Partial Invariance Model \citep{byrneTestingEquivalenceFactor1989}:

\begin{enumerate}
    \item Identifying Sources of Non-invariance: We examine the Critical Ratios for Differences between parameters. A Z-score with an absolute value exceeding 1.96 indicates significant inter-group heterogeneity at the 0.05 significance level. This diagnostic approach enables the precise identification of specific parameters that contribute to the deterioration of model fit.
    \item Iterative Release of Constraints: Following a "Stepwise Release" strategy, parameters exhibiting the highest significant discrepancies are sequentially freed across groups. This process continues until the $\Delta \chi^2$ between the modified model and the nested model becomes non-significant, ensuring that the model's fit is no longer compromised by inappropriate equality constraints.
    \item Establishment of Partial Invariance: The final resulting model is recognized as a Partial Invariance Model. This procedure allows for a granular understanding of which specific structural paths vary between the Student Group and the Professional Group while maintaining the validity of cross-group comparisons \citep{steenkampAssessingMeasurementInvariance1998}.
\end{enumerate}

\section{Results}
\subsection{Data Exploration}
\begin{figure}[htbp]
	\centering
		\includegraphics[scale=0.325]{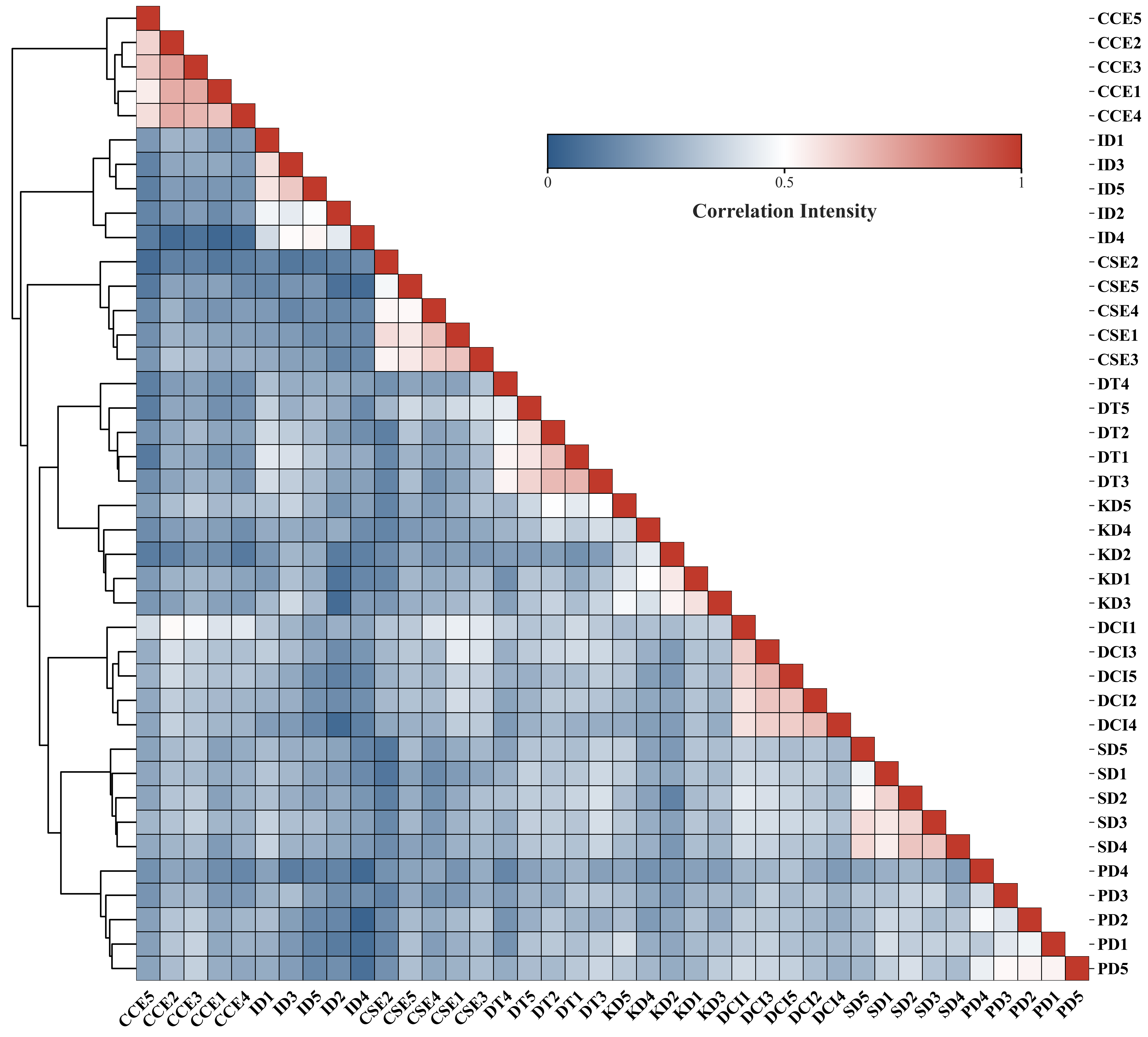}
	\caption{Hierarchical Clustered Correlation Heatmap for Exploratory Data Analysis}
	\label{fig: Correlation}
\end{figure}

The hierarchical clustering correlation heatmap is presented in Figure \ref{fig: Correlation}. Exploratory analysis of the correlation relationships reveals a pronounced diagonal block structure, where items belonging to the same construct form distinctive high-intensity clusters. These observed items are consistently partitioned into identical branches of the clustering dendrogram and exhibit strong positive correlations, providing robust evidence for convergent validity. Simultaneously, the heatmap highlights a distribution pattern characterized by “high within-group similarity and low between-group correlation”; regions corresponding to heterogeneous constructs show noticeably lighter color intensity, indicating relatively low correlation coefficients. This well-delineated color gradient offers intuitive and compelling empirical support for discriminant validity.

\begin{figure}[h]
	\centering
		\includegraphics[scale=.3]{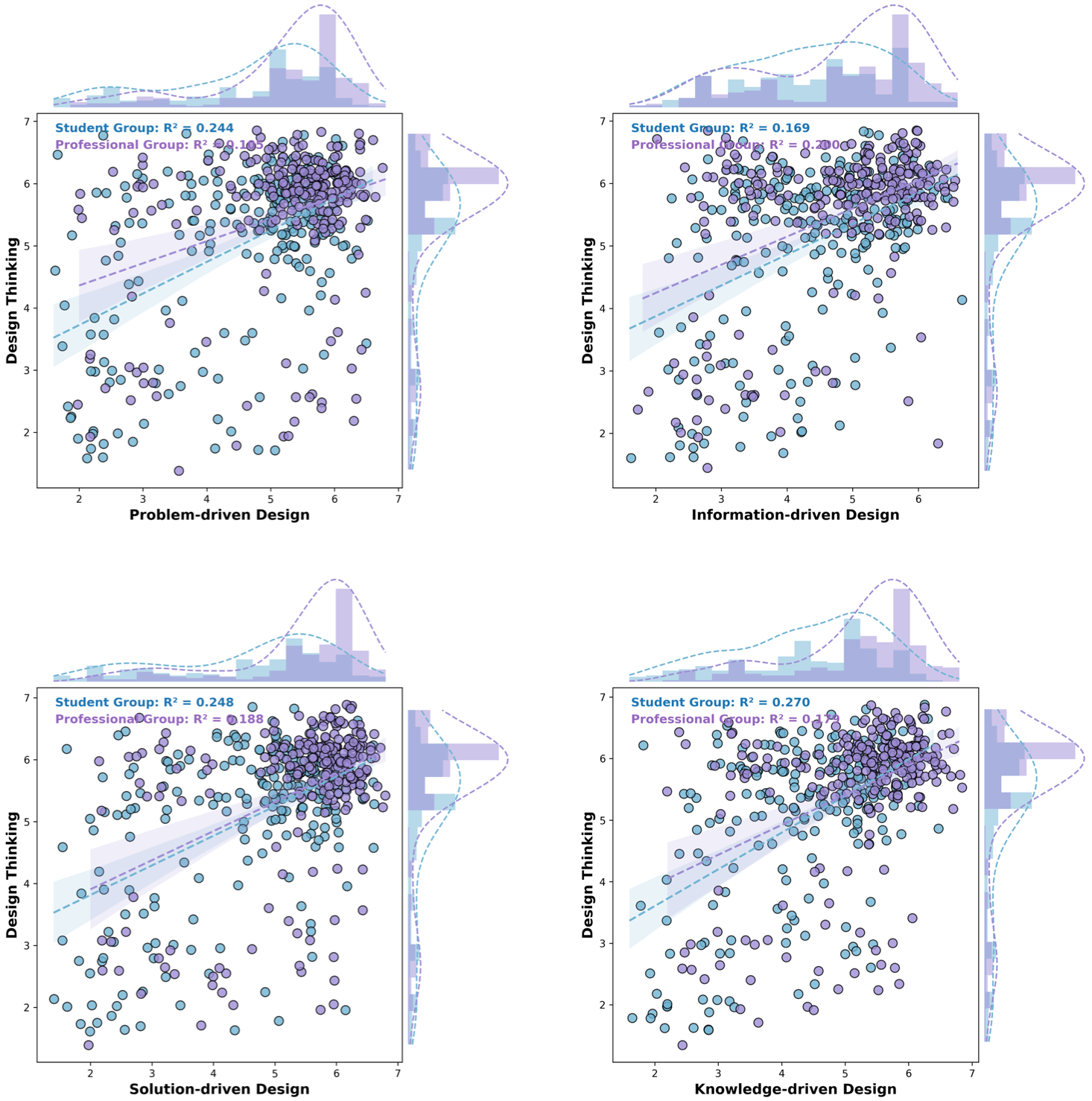}
	\caption{Comparative Regression Analysis of Four Thinking Skills on Design Thinking Across Experience Levels}
	\label{fig: Comparative Regression Analysis}
\end{figure}

As shown in Figure \ref{fig: Comparative Regression Analysis}, this study employed grouped linear regression and joint distribution analysis to examine the differential effects of Problem-driven Design, Information-driven Design, Solution-driven Design, and Knowledge-driven Design on Design Thinking. The regression results, as shown in Table \ref{tab:regression_analysis}, indicate that all models across the tested dimensions exhibit a high level of statistical significance ($p < 0.001$). The F-statistics of the regression models range from 28.597 to 84.348, demonstrating that Thinking Skills serve as robust and effective predictors of Design Thinking. In the Problem-driven Design dimension, the student group achieved a coefficient of determination of $R^2 = 0.244$, with a regression slope of $\beta = 0.508$ and an intercept of 2.707; by comparison, the corresponding values for the professional group were $R^2 = 0.105$, $\beta = 0.356$, and an intercept of 3.647. In the Information-driven Design dimension, the regression results for the student group yielded $R^2 = 0.169$, $\beta = 0.494$, and an intercept of 2.887. In contrast, the professional group exhibited higher explanatory power in this dimension, with $R^2 = 0.200$, a regression slope of $\beta = 0.451$, and an intercept of 3.345. In the Solution-driven Design dimension, the student group demonstrated relatively strong explanatory power ($R^2 = 0.248$), with a regression slope of $\beta = 0.478$ and an intercept of 2.858. The professional group showed the closest slope to that of the student group ($\beta = 0.470$), while its coefficient of determination was $R^2 = 0.188$ and the intercept was 2.966. Finally, in the Knowledge-driven Design dimension, the regression analysis results for the student group were $R^2 = 0.270$, $\beta = 0.596$, and an intercept of 2.412, while for the professional group, the results were $R^2 = 0.179$, $\beta = 0.486$, and an intercept of 2.979.

Overall, the comparative results indicate that across all observed dimensions, the regression slopes for the student group are consistently higher than those for the professional group, and the student group also demonstrates clear advantages in explanatory power in the Problem-driven Design, Solution-driven Design, and Knowledge-driven Design dimensions. In addition, the intercepts across all regression models are consistently higher for the professional group, suggesting a higher intrinsic baseline of Design Thinking.

\begin{figure}
	\centering
		\includegraphics[scale=.425]{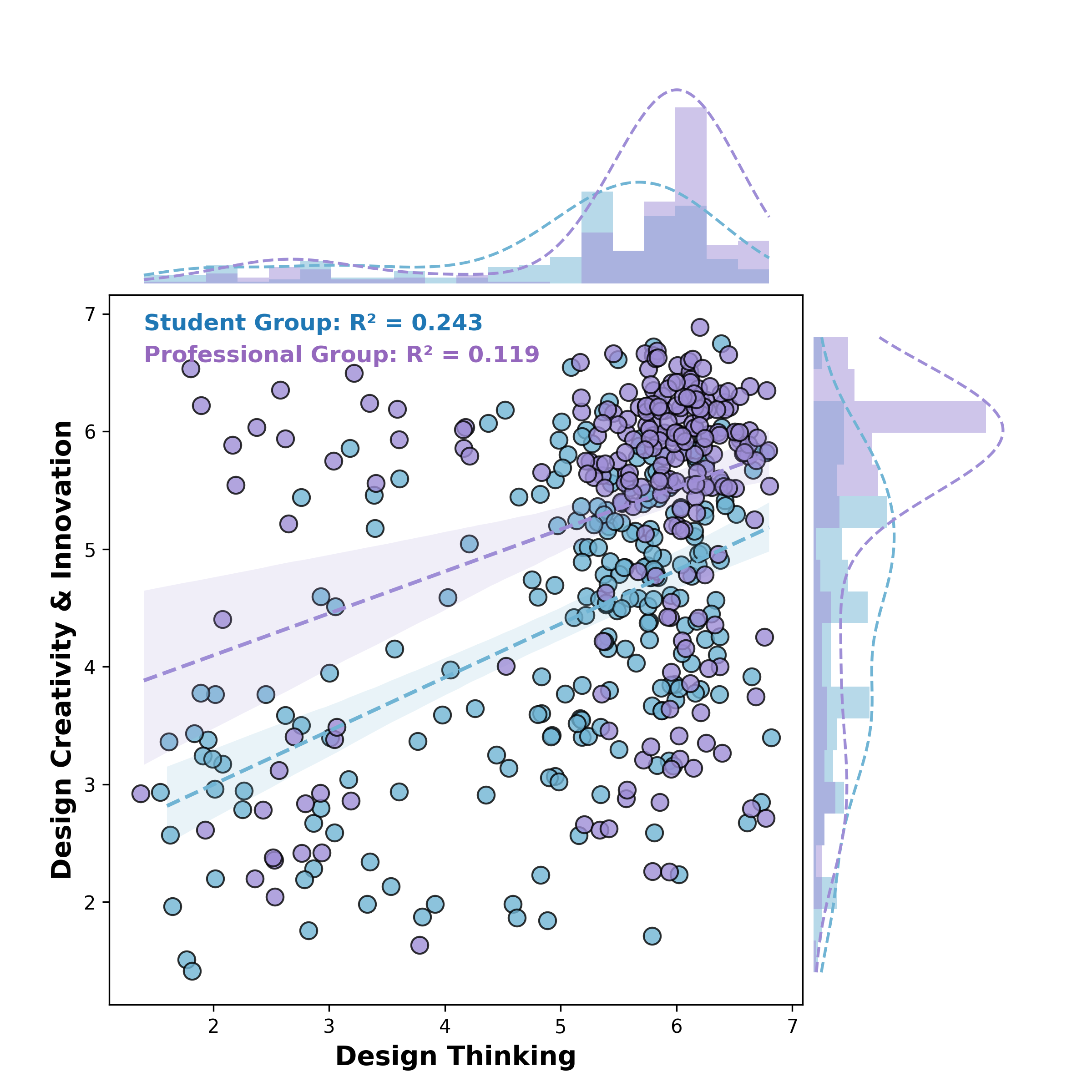}
	\caption{Comparative Regression Analysis of Design Thinking on Design Creativity \& Innovation Across Experience Levels}
	\label{fig: Design Thinking on Design Creativity}
\end{figure}

The results of the grouped linear regression and joint distribution analysis of Design Thinking on Design Creativity \& Innovation are shown in Figure \ref{fig: Design Thinking on Design Creativity}. Specifically, as shown in Table \ref{tab:regression_analysis}, the regression model for the student group achieved a coefficient of determination of $R^2 = 0.243$, a regression slope of $\beta = 0.455$, and an intercept of 2.087; by comparison, the regression results for the professional group yielded $R^2 = 0.119$, a regression slope of $\beta = 0.357$, and an intercept of 3.382. 

The comparative analysis shows that the student group outperforms the professional group in terms of regression slope and explanatory power, indicating that student creativity is more sensitive to the application of Design Thinking methodologies and significantly dependent on the systematic nature of these methods. At the same time, the professional group exhibits a higher baseline level of innovation, indicating that the professional group already possesses strong innovation capabilities without external intervention. Furthermore, the marginal distribution plot demonstrates that the professional group's sample points are more significantly concentrated, with a density curve peak notably higher than that of the student group, suggesting higher consistency and stability in professional outputs; in contrast, the wider distribution of the student group reflects greater individual variance and a more significant range for potential growth.

\begin{table}[htbp]
\centering
\caption{Summary of Linear Regression Analysis Results}
\label{tab:regression_analysis}
\begin{tabular*}{\linewidth}{@{\extracolsep{\fill}}llccccccc@{}}
\toprule
Model & Group & $R^2$ & Adj.$R^2$ & Coef. ($\beta$) & Intercept & S.E. & F & Sig. \\
\midrule
PD $\rightarrow$ DT & Student & 0.244 & 0.241 & 0.508 & 2.707 & 0.059 & 73.538 & *** \\
                    & Professional & 0.105 & 0.102 & 0.356 & 3.647 & 0.067 & 28.597 & *** \\
\addlinespace
ID $\rightarrow$ DT & Student & 0.169 & 0.166 & 0.494 & 2.887 & 0.072 & 46.466 & *** \\
                    & Professional & 0.200 & 0.197 & 0.451 & 3.345 & 0.058 & 60.787 & *** \\
\addlinespace
SD $\rightarrow$ DT & Student & 0.248 & 0.244 & 0.478 & 2.858 & 0.055 & 75.075 & *** \\
                    & Professional & 0.188 & 0.185 & 0.470 & 2.966 & 0.063 & 56.244 & *** \\
\addlinespace
KD $\rightarrow$ DT & Student & 0.270 & 0.267 & 0.596 & 2.412 & 0.065 & 84.348 & *** \\
                    & Professional & 0.179 & 0.176 & 0.486 & 2.979 & 0.067 & 53.104 & *** \\
\addlinespace
DT $\rightarrow$ DCI & Student & 0.243 & 0.240 & 0.455 & 2.087 & 0.053 & 73.176 & *** \\
                    & Professional & 0.119 & 0.115 & 0.357 & 3.382 & 0.062 & 32.808 & *** \\
\bottomrule
\end{tabular*}
\end{table}

\subsection{Structural Equation Model}
We conducted a confirmatory factor analysis and constructed a structural equation model based on the theoretical framework. The model fit results are presented in Table \ref{tab: model fit}. The results show that $\chi^2/\text{df}$ is 1.866, CFI is 0.937, TLI is 0.932, RMSEA is 0.043, and SRMR is 0.078, all of which meet the good fit criteria. Although GFI is 0.877 and AGFI is 0.861, , which did not reach the ideal level, they all fall within an acceptable range. Considering the complexity and exploratory nature of research in the design discipline, perfect fit results are relatively rare \citep{marshSearchGoldenRules2004}. Therefore, we consider the overall fit of the model to be good, which sufficiently validates the rationality of the research model \citep{dollConfirmatoryFactorAnalysis1994}.

\begin{table}[h]
\caption{The Evaluation Indicators and Assessment of Model Fit in Structural Equation Modeling}
\label{tab: model fit}
\begin{tabular*}{\linewidth}{@{\extracolsep{\fill}}lccccccc@{}}
\toprule
          & $\chi^2/\text{df}$ & GFI   & AGFI  & CFI   & TLI   & RMSEA & SRMR  \\
\midrule
Cutoff value    & $<$ 3.00  & $>$ 0.900& $>$ 0.900& $>$ 0.900& $>$ 0.900& $<$ 0.080 & $<$ 0.080 \\
Analysis Results& 1.866     & 0.877    & 0.861    & 0.937    & 0.932    & 0.043     & 0.078     \\
Model Evaluation& Good fit  & Accept   & Accept   & Good fit & Good fit & Good fit  & Good fit  \\
\bottomrule
\end{tabular*}
\end{table}

Table \ref{tab: measurement_model} in the Appendix summarizes the factor loadings, composite reliability, and average variance extracted of the measurement model. The results indicate that all standardized factor loadings exceed the 0.500 threshold, confirming sufficient indicator reliability for each construct. Regarding internal consistency, the composite reliability for all latent variables surpasses 0.800, markedly exceeding the generally accepted criterion of 0.700, which demonstrates high reliability. In terms of convergent validity, the average variance extracted for most constructs satisfies the 0.500 benchmark. Notably, although the average variance extracted values for "Problem-driven Design" and "Knowledge-driven Design" are slightly below 0.500. According to the conservative criteria established by Fornell and Larcker \citep{fornellEvaluatingStructuralEquation1981}, the convergent validity of a construct is considered adequate if its average variance extracted is below 0.500, provided that its composite reliability is higher than 0.600. Given that the composite reliability values in this study far exceed both the 0.600 thresholds, the measurement model demonstrates robust reliability and satisfactory convergent validity across all dimensions.

Table \ref{tab: path coefficients} summarizes the estimated path coefficients of the structural equation model. The results indicate that all hypothesized paths reached statistical significance ($p < 0.001$). At the level of Thinking Skills, Problem-driven Design ($\beta = 0.198$, $p < 0.01$), Information-driven Design ($\beta = 0.241$, $p < 0.001$), Solution-driven Design ($\beta = 0.227$, $p < 0.001$), and Knowledge-driven Design ($\beta = 0.263$, $p < 0.001$) all exerted significant positive effects on Design Thinking. As a core cognitive driving force, Design Thinking demonstrated strong predictive power, significantly enhancing both Creative Self-Efficacy ($\beta = 0.462$, $p < 0.001$) and Collective Creative Efficacy ($\beta = 0.280$, $p < 0.001$). In addition, Design Thinking maintained a direct and significant positive effect on Design Creativity \& Innovation ($\beta = 0.286$, $p < 0.001$). Furthermore, the internal psychological mechanisms revealed by the model were also supported, with Creative Self-Efficacy exerting a significant positive influence on Collective Creative Efficacy ($\beta = 0.211$, $p < 0.001$). At the outcome level, both efficacy constructs emerged as important predictors of Design Creativity \& Innovation performance. Specifically, Creative Self-Efficacy ($\beta = 0.331$, $p < 0.001$) and Collective Creative Efficacy ($\beta = 0.309$, $p < 0.001$) significantly enhanced Design Creativity \& Innovation.

\begin{table}[htbp]
\caption{Estimation of Path Coefficients in Structural Equation Modeling}
\label{tab: path coefficients}
\begin{tabular*}{\linewidth}{@{\extracolsep{\fill}}lclccccc@{}}
\toprule
Path Analyses                 &                 &                                 & Estimate & S.E.  & t-value & Sig.    & Std.  \\
\midrule
Problem-driven Design         & $\rightarrow$   & Design thinking                 & 0.250    & 0.078 & 3.213   & **      & 0.198 \\
Information-driven Design     & $\rightarrow$   & Design thinking                 & 0.289    & 0.061 & 4.728   & ***     & 0.241 \\
Solution-driven Design        & $\rightarrow$   & Design thinking                 & 0.295    & 0.077 & 3.852   & ***     & 0.227 \\
Knowledge-driven Design       & $\rightarrow$   & Design thinking                 & 0.290    & 0.065 & 4.433   & ***     & 0.263 \\
Design thinking               & $\rightarrow$   & Creative Self-Efficacy          & 0.448    & 0.050 & 8.997   & ***     & 0.462 \\
Design thinking               & $\rightarrow$   & Collective Creative Efficacy    & 0.280    & 0.056 & 5.013   & ***     & 0.280 \\
Creative Self-Efficacy        & $\rightarrow$   & Collective Creative Efficacy    & 0.217    & 0.057 & 3.789   & ***     & 0.211 \\
Design thinking               & $\rightarrow$   & Design Creativity \& Innovation & 0.240    & 0.041 & 5.843   & ***     & 0.286 \\
Creative Self-Efficacy        & $\rightarrow$   & Design Creativity \& Innovation  & 0.286    & 0.042 & 6.740   & ***     & 0.331 \\
Collective Creative Efficacy  & $\rightarrow$   & Design Creativity \& Innovation  & 0.260    & 0.038 & 6.784   & ***     & 0.309 \\
\bottomrule
\end{tabular*}
\end{table}

\subsection{Mediation Analysis}
Table~\ref{tab: mediation effect} presents the mediation analysis results based on 5,000 bootstrap samples. The indirect effect of Creative Self-Efficacy was $0.128^{***}$ ($95\%$ Percentile CI $[0.079, 0.188]$, BC CI $[0.081, 0.191]$). Similarly, the indirect effect of Collective Creative Efficacy was $0.073^{**}$ ($95\%$ Percentile CI $[0.037, 0.119]$, BC CI $[0.038, 0.121]$). Furthermore, the chain mediation path (CSE $\rightarrow$ CCE) yielded a point estimate of $0.025^{**}$ ($95\%$ Percentile CI $[0.010, 0.045]$, BC CI $[0.011, 0.048]$). Since none of the $95\%$ confidence intervals include zero, these results provide strong evidence for the significant parallel and chain mediation effects of Creative Self-Efficacy and Collective Creative Efficacy.

Additionally, a contrast analysis was conducted to compare the magnitudes of different indirect paths. The contrast between the indirect effects of Creative Self-Efficacy and Collective Creative Efficacy was not statistically significant (point estimate = $0.055$, $95\%$ Percentile CI $[-0.014, 0.129]$, BC CI $[-0.011, 0.132]$), as both intervals included zero. However, the indirect effects of both Creative Self-Efficacy and Collective Creative Efficacy were significantly stronger than the chain mediation effect (CSE $\rightarrow$ CCE). Specifically, the contrast estimate for Creative Self-Efficacy vs. the chain path was $0.103^{***}$ ($95\%$ Percentile CI $[0.054, 0.162]$, BC CI $[0.057, 0.164]$), and for Collective Creative Efficacy vs. the chain path was $0.047^{*}$ ($95\%$ Percentile CI $[0.007, 0.097]$, BC CI $[0.008, 0.098]$). These results indicate that while the two parallel mediators have comparable strengths, their individual indirect effects significantly exceed their combined chain effect.

To further quantify the explanatory weight of these mechanisms, effect proportions and their respective 95\% bootstrap confidence intervals were calculated. The results indicate that the total indirect effect constitutes $48.5\%^{***}$ of the total effect ($95\%$ Percentile CI $[0.368, 0.625]$; BC CI $[0.369, 0.626]$). Specifically, the contributions of Creative Self-Efficacy, Collective Creative Efficacy, and the chain mediation path to the total effect are $27.5\%^{***}$ ($95\%$ Percentile CI $[0.180, 0.397]$; BC CI $[0.183, 0.406]$), $15.6\%^{***}$ ($95\%$ Percentile CI $[0.084, 0.245]$; BC CI $[0.087, 0.248]$), and $5.4\%^{**}$ ($95\%$ Percentile CI $[0.021, 0.097]$; BC CI $[0.023, 0.101]$), respectively. Furthermore, when examining the internal composition of the total indirect effect, Creative Self-Efficacy emerges as the primary driver, accounting for $56.7\%^{***}$ of the total indirect effect ($95\%$ Percentile CI $[0.420, 0.712]$; BC CI $[0.418, 0.711]$). This is followed by Collective Creative Efficacy, which represents $32.1\%^{***}$ ($95\%$ Percentile CI $[0.174, 0.475]$; BC CI $[0.176, 0.476]$), while the chain mediation path accounts for $11.2\%^{**}$ ($95\%$ Percentile CI $[0.047, 0.186]$; BC CI $[0.051, 0.192]$). Since all $95\%$ confidence intervals for these proportion estimates exclude zero, these findings confirm the robust and significant explanatory contribution of each mediation path within the overall model.

\begin{table}[htbp]
\centering
\caption{Mediation Effect Analysis Based on Bootstrapping Method}
\label{tab: mediation effect}
\begin{tabular*}{\linewidth}{@{\extracolsep{\fill}}
    l 
    S[table-format=1.3]   
    c                     
    S[table-format=1.3]   
    S[table-format=1.3]   
    S[table-format=-1.3]  
    S[table-format=1.3]   
    S[table-format=-1.3]  
    S[table-format=1.3]   
@{}}
\toprule
& \multicolumn{1}{c}{\multirow{3}{*}{\makecell{Point\\Estimate}}} 
& \multicolumn{1}{c}{\multirow{3}{*}{\makecell{Sig.}}}
& \multicolumn{2}{c}{\makecell{Product of\\Coefficients}} 
& \multicolumn{4}{c}{Bootstrapping (95\% CI)} \\
\cmidrule(lr){4-5} \cmidrule(lr){6-9}
& 
&
& \multicolumn{1}{c}{\multirow{2}{*}{SE}} 
& \multicolumn{1}{c}{\multirow{2}{*}{Z}} 
& \multicolumn{2}{c}{Percentile} 
& \multicolumn{2}{c}{Bias-Corrected} \\
\cmidrule(lr){6-7} \cmidrule(lr){8-9}
& & & & & \multicolumn{1}{c}{Lower} & \multicolumn{1}{c}{Upper} & \multicolumn{1}{c}{Lower} & \multicolumn{1}{c}{Upper} \\
\midrule
\multicolumn{9}{c}{Indirect Effects} \\
\midrule
CSE                           & 0.128 & *** & 0.028 & 4.596 &  0.079 & 0.188 &  0.081 & 0.191 \\
CCE                           & 0.073 & **  & 0.021 & 3.425 &  0.037 & 0.119 &  0.038 & 0.121 \\
CSE $\rightarrow$ CCE         & 0.025 & **  & 0.009 & 2.860 &  0.010 & 0.045 &  0.011 & 0.048 \\
Total Indirect                & 0.226 & *** & 0.037 & 6.170 &  0.160 & 0.302 &  0.161 & 0.314 \\
\midrule
\multicolumn{9}{c}{Total Effect} \\
\midrule
Total Effect                  & 0.466 & *** & 0.058 & 8.001 &  0.356 & 0.587 &  0.355 & 0.586 \\
\midrule
\multicolumn{9}{c}{Contrasts} \\
\midrule
CSE vs. CCE                   & 0.055 &     & 0.036 & 1.541 & -0.014 & 0.129 & -0.011 & 0.132 \\
CSE vs. CSE $\rightarrow$ CCE & 0.103 & *** & 0.027 & 3.762 &  0.054 & 0.162 &  0.057 & 0.164 \\
CCE vs. CSE $\rightarrow$ CCE & 0.047 & *   & 0.023 & 2.059 &  0.007 & 0.097 &  0.008 & 0.098 \\
\bottomrule
\end{tabular*}
\end{table}

\subsection{Multi-Group Analysis}
To evaluate the structural invariance of the proposed model across groups with varying levels of design experience, a multi-group analysis was conducted. This analysis sequentially tested for configural, measurement weights, structural weights, and structural covariances invariance. The results are summarized in Table \ref{tab:invariance_results}.

\begin{enumerate}
    \item Configural Invariance (M1): The configural invariance model (M1), which imposes no equality constraints across groups, served as the baseline for nested comparisons. The results indicated that M1 exhibited an acceptable fit ($\chi^2 = 2256.6, df = 1448, CFI = 0.911, RMSEA = 0.048$), suggesting that the basic model structure is consistent across both the student and professional groups.
    
    \item Measurement Invariance (M2):  The full measurement invariance model (M2-Full), where all factor loadings were constrained to be equal, showed a significant difference compared to M1 ($\Delta\chi^2(32) = 51.031, p = 0.018$), indicating heterogeneity in factor loadings. To identify the sources of non-invariance, we examined the Critical Ratios for Difference (Z-values) in Table \ref{tab: item_weight_comparison}. Following an iterative modification process, constraints were sequentially released based on the descending order of their Z-values. Releasing the constraint on DCI3 ($Z = 3.289, p < 0.01$) was insufficient to achieve model equivalence; subsequently, the constraint on DCI4 ($Z = 2.916, p < 0.01$) was also released. After relaxing these two parameters, the partial measurement model (M2-P) showed no significant difference compared to M1 ($\Delta\chi^2(30) = 42.437, p = 0.066$), thus establishing partial measurement invariance. 

    \item Structural Weight Invariance (M3): Based on the M2-P model, the structural weight invariance test was performed. The full weight model (M3-Full) showed a significant chi-square difference relative to M2-P ($\Delta\chi^2(10) = 19.049, p = 0.040$). Table \ref{tab: structural_path_comparison} revealed that the path from Creative Self-Efficacy to Collective Creative Efficacy exhibited the most significant discrepancy between groups ($Z = -2.899, p < 0.01$). After releasing the constraint on this specific path, the partial structural weight model (M3-P) showed no significant difference compared to M2-P ($\Delta\chi^2(9) = 14.950, p = 0.092$), confirming that the remaining structural path coefficients were invariant across groups.
    
    \item Structural Covariance Invariance (M4): Finally, we examined the invariance of structural variances and covariances. The full covariance model (M4-Full) showed a significant difference compared to M3-P ($\Delta\chi^2(10) = 26.796, p = 0.003$). Table \ref{tab: variance_covariance_comparison} indicated heterogeneity among several exogenous constructs. To identify the most parsimonious set of adjustments, we performed a series of comparisons and found that releasing three specific constraints, specifically the variances of Problem-driven Design and Solution-driven Design as well as the covariance between Information-driven Design and Knowledge-driven Design, was sufficient to achieve model equivalence. The resulting partial covariance model (M4-P) showed no significant difference relative to M3-P ($\Delta\chi^2(7) = 13.644, p = 0.058$), indicating that the remaining structural variances and covariances were consistent across groups.
\end{enumerate}

Overall, the establishment of partial invariance confirms that the core structural relationships of the proposed model are largely consistent across groups, allowing for meaningful comparisons. Beyond confirming generalizability, the results provide critical empirical evidence for domain-specific applications. Specifically, the observed differences in efficacy transformation mechanisms and design drivers highlight the distinct psychological and operational profiles of students versus professionals, informing more precise pedagogical and organizational strategies.

\begin{table}[htbp]
\centering
\caption{Summary of Multi-Group Analysis Results}
\label{tab:invariance_results}
\begin{tabular*}{\textwidth}{@{\extracolsep{\fill}} 
    l                   
    S[table-format=4.1]  
    S[table-format=4.0]  
    S[table-format=1.3]  
    S[table-format=1.3]  
    c                   
    S[table-format=2.3]  
    S[table-format=2.0]  
    S[table-format=1.3{**}] 
    S[table-format=-1.3] 
    c                   
@{}}
\toprule
{Model} & {$\chi^2$} & {$df$} & {CFI} & {RMSEA} & {Comp.} & {$\Delta\chi^2$} & {$\Delta df$} & {$p$} & {$\Delta$CFI} & {Decision} \\
\midrule
M1 & 2256.6 & 1448 & 0.911 & 0.048 & {--} & {--} & {--} & {--} & {--} & {--} \\
\addlinespace

M2-Full & 2307.7 & 1480 & 0.908 & 0.049 & M1 & 51.031 & 32 & 0.018\textsuperscript{*} & -0.002 & Reject \\
M2-P & 2299.1 & 1478 & 0.909 & 0.048 & M1 & 42.437 & 30 & 0.066 & -0.001 & Accept \\
\multicolumn{11}{l}{\quad \footnotesize (Partial: DCI $= \sim$ DCI3, DCI $= \sim$ DCI4)} \\
\addlinespace

M3-Full & 2318.1 & 1488 & 0.908 & 0.048 & M2-P & 19.049 & 10 & 0.040\textsuperscript{*} & -0.001 & Reject \\
M3-P & 2314.0 & 1487 & 0.908 & 0.048 & M2-P & 14.950 & 9 & 0.092 & -0.001 & Accept \\
\multicolumn{11}{l}{\quad \footnotesize (Partial: CCE $\sim$ CSE)} \\
\addlinespace

M4-Full & 2340.8 & 1497 & 0.907 & 0.049 & M3-P & 26.796 & 10 & 0.003\textsuperscript{**} & -0.002 & Reject \\
M4-P & 2327.7 & 1494 & 0.908 & 0.048 & M3-P & 13.644 & 7 & 0.058 & -0.001 & Accept \\
\multicolumn{11}{l}{\quad \footnotesize (Partial: PD $\sim\sim$ PD, ID $\sim\sim$ KD, SD $\sim\sim$ SD)} \\
\bottomrule
\end{tabular*}
\end{table}

\section{Discussion}
\subsection{Linear Regression and Structural Equation Modeling Cross-validation}
Among the four Thinking Skills influencing Design Thinking, both regression analysis and structural equation models demonstrate that Knowledge-driven Design has the most significant impact. This suggests that Design Creativity \& Innovation is primarily driven by systematic knowledge reserves and the integration of experience, rather than by transient inspiration \citep{crossDesignerlyWaysKnowing2001, dorstCoreDesignThinking2011}. Specifically, a robust knowledge base provides the necessary semantic framework for designers to move beyond superficial imitations and grasp the underlying logical structures of a problem, enabling the synthesis of complex information into coherent design strategies. Consequently, the impetus for innovation stems from the ability to retrieve and reconfigure domain-specific insights, which effectively transforms the inherent uncertainty of the design process into a controlled and strategic exploration.

Notably, multi-group analysis reveals a structural inflection in the "Solution-driven" dimension. While students generally exhibit higher path coefficients across other dimensions, the "Solution-driven" path yields higher coefficients and greater robustness within the professional group. This phenomenon signifies a paradigmatic shift in cognitive strategy: while novices tend to be trapped in problem-scoping, experts employ solutions as epistemic probes to navigate the co-evolution of problem and solution spaces \citep{dorstCreativityDesignProcess2001}. For instance, the emphasis on using ideas to guide information search (SD1) and maintaining a diverse pool of feasible solutions (SD2, SD5) reflects the expert strategy of using "solution conjectures" to simultaneously define and solve the problem. As Dorst and Cross \citep{dorstCreativityDesignProcess2001} noted, creative design involves a constant oscillation between problem and solution spaces; thus, the professional group’s efficiency in generating multiple ideas (SD3, SD4) functions as a high-frequency cognitive mechanism to navigate this co-evolution. For professionals, a solution is no longer a mere terminal output but a cognitive vehicle for deconstructing ill-defined problems and synthesizing tacit knowledge through iterative reflection-in-action \citep{schonReflectivePractitionerHow1983}.

Furthermore, the linear regression results reveal the pedagogical logic: the higher slope exhibited by the student group indicates a golden window for instructional intervention, whereas the high intercept in the professional group reflects a robust cognitive baseline established through long-term professional accumulation \citep{dreyfusMindMachinePower1988}. Consequently, the goal of design education is not merely to accelerate growth but to elevate the cognitive baseline of students through sustained training, effectively transforming design thinking from a conscious tool into an internalized professional instinct.

Finally, this study elucidates the core essence of design professionalism, which is defined by robust and predictable output quality rather than fleeting inspiration \citep{ericssonCambridgeHandbookExpertise2013}. Joint Distribution Analysis reveals a stark contrast between the high consistency of the professional group and the high volatility of the student group, mirroring a staged evolution of thought from spontaneous sensibility to conscious rationality \citep{crossDesignThinkingUnderstanding2016}. Therefore, design education should establish a pedagogical model grounded in knowledge, driven by efficacy, and centered on stability, effectively guiding students to transition from volatile emotional creation toward high-baseline professional delivery.

\subsection{Structural Equation Modeling and Mediation Effects}
From the significance of psychological empowerment, this study reveals no significant difference in the magnitude of the mediating effects between Creative Self-Efficacy and Collective Creative Efficacy. This balanced empirical result indicates that in the innovation process, an individual’s belief in their creative capabilities is as vital as their confidence in team collaboration \citep{stajkovicCollectiveEfficacyGroup2009}. Design Thinking does not merely favor individual creativity or collective cooperation; instead, it facilitates a dynamic equilibrium between "individual deep thinking" and the "emergence of collective intelligence." This model refutes the bias toward either "individual heroism" or "pure teamwork," demonstrating that superior innovation output is predicated on the co-evolution of both individual and collective beliefs.

More critically, the structural equation modeling results successfully deconstruct the psychological black box of design education. With the total indirect effect accounting for approximately 48.5\% of the total effect, this research provides robust empirical evidence that psychological empowerment holds equal strategic value to methodological instruction \citep{tierneyCreativeSelfEfficacyIts2002, banduraSelfefficacyExerciseControl2012}. These findings suggest that innovation output is essentially a balanced system sustained by both rational logic tools and emotional psychological beliefs. This insight establishes a critical demarcation for practical application. If the implementation of Design Thinking remains at a superficial level characterized by sticky notes, sketches, and procedural compliance while neglecting the activation of underlying psychological pathways, the innovative potential of methodological training will encounter a severe conversion bottleneck. Consequently, future practices must shift from being process-driven to cognition-driven, ensuring a profound resonance between the hardware of methodical tools and the software of psychological cognition. 

Furthermore, although the absolute effect size of the chain mediation is smaller than those of the independent paths, it establishes a critical logical sequence: Creative Self-Efficacy serves as a precursor to Collective Creative Efficacy. This implies that the elevation of collective efficacy is not an arbitrary phenomenon but is firmly rooted in the awakening of each member's self-efficacy. These findings suggest that Design Thinking improves interpersonal interactions within the team by enhancing Creative Self-Efficacy, thereby boosting the overall team effectiveness. For managers and educators, this provides a clear intervention strategy: the breakthrough for enhancing team performance often lies in the prior empowerment of the individual.

\subsection{Differentiating Multi-Group Comparisons from a Design Education Perspective}
First, the failure of measurement invariance reveals a fundamental cognitive disconnect between the two groups regarding the definition of Design Creativity \& Innovation, as partial invariance (M2-P) was attained only after relaxing the constraints on DCI3 and DCI4. This data nuance exposes a critical bias in contemporary design education: students tend to over-identify innovation quality with procedural intensity, equating superior design with the "accumulation of divergent ideas" (DCI3) and the "frequency of logical adjustments" (DCI4). Conversely, the professional group’s lower scores on these procedural indicators suggest a transition from "explicit effort" to "internalized expertise." For professionals, innovation is not a product of repetitive procedural operations but an outcome of high-precision decision-making. This divergence implies that pedagogical focus should shift from merely encouraging "procedural busyness" toward fostering "strategic parsimony," guiding students to transition from novice "labor-intensive trials" toward the selective optimization that characterizes professional maturity.

Secondly, the significant disparities in path coefficients expose a structural decoupling between individual confidence and collective synergy within the student group \citep{banduraExerciseHumanAgency2000}. Achieving partial invariance by relaxing the constraints on the "CCE $\rightarrow$ CSE" path uncovers a fundamental rift in how efficacy is transmitted between the two cohorts. Specifically, the standardized path coefficient from Creative Self-Efficacy to Collective Creative Efficacy is 0.314 in the professional group, while it plummets to a mere 0.079 in the student group. This near-total decoupling suggests that in academic environments, "confident individuals" do not inherently coalesce into "confident teams." This serves as a vital wake-up call for design education: if pedagogical focus remains confined to individual skill sets while neglecting the "Relational Design" of collaborative dynamics, personal talents will persist as isolated islands incapable of generating synergy. Consequently, educators must implement structured team-coordination interventions, including the cultivation of psychological safety and the establishment of mutual trust protocols, to bridge this functional chasm between the individual and the collective.

Furthermore, the rejection of the structural covariance invariance model confirms fundamental heterogeneities in the cognitive structures. To achieve partial invariance (M4-P), the variances of Problem-driven Design and Solution-driven Design, along with the "Information-driven Design $\leftrightarrow$ Knowledge-driven Design" covariance, had to be freed, precisely identifying the logical divergence points in Design Thinking. The results highlight two core cognitive bottlenecks for the student group. First, compared to professionals, students' cognitive mechanism lacks "anchoring effects." The higher variances in Problem-driven Design (1.294) and Solution-driven Design (1.020) relative to the professional group reflect distinct non-steady-state characteristics and high cognitive volatility \citep{crossExpertiseDesignOverview2004}. Due to the absence of a stable logical starting point compared to their professional counterparts, students struggle to achieve cognitive focus when facing complex design tasks. Second, there is a "knowledge integration gap." The correlation between Information-driven Design and Knowledge-driven Design in students is markedly weaker than the robust synergy exhibited by professionals. This indicates that gathered information remains superficial for students and is not effectively internalized into actionable design knowledge, leading to a lack of deep logical support between information acquisition and design decision-making.

Finally, the systematic rejection of full structural invariance (from M2 to M4) establishes an empirical foundation for adaptive pedagogy in design education. The results demonstrate that the underlying logic of Design Thinking operates through distinct structural configurations within academic and professional domains. This non-invariance should be interpreted not as a pedagogical failure, but rather as a Targeted Intervention Map. Educators should refrain from forcing students to mechanically adopt professional standards for which they lack the psychological foundation; instead, they should utilize these findings to reinforce specific weak links. This involves guiding students to ground their creative solutions in systematic knowledge bases in order to resolve the "knowledge-execution disconnect," while also facilitating the arduous transition from individual confidence toward collective synergy.

\subsection{Research Limitations}
This study is subject to several limitations. First, although design is recognized as an inherently practice-oriented discipline \citep{crossDesignThinkingUnderstanding2016}, the assessment of Design Creativity \& Innovation in this research relies primarily on survey data, thereby lacking a direct evaluation of concrete design outputs \citep{johansson-skoldbergDesignThinkingPresent2013}. To address this, future research should incorporate empirical case and qualitative assessments of design products to more comprehensively validate the efficacy of Design Thinking in practical environments. Second, the heavy reliance on self-reported measures may be susceptible to cognitive biases \citep{dunningFlawedSelfAssessmentImplications2004}. Future studies would benefit from incorporating third-party objective assessments or employing data triangulation from multiple sources to further bolster the reliability and validity of the findings.

\section{Conclusions}
This study establishes that Design Thinking, defined through the integration of Problem-driven, Information-driven, Solution-driven, and Knowledge-driven Design, serves as the fundamental cornerstone of Design Creativity \& Innovation. Our findings reveal a dual mechanism: Design Thinking not only exerts a significant direct influence on innovation outcomes but also facilitates psychological empowerment by enhancing Creative Self-Efficacy and Collective Creative Efficacy. This impact underscores that technical methodology and psychological empowerment are equally indispensable within the creative process. A pivotal discovery of this research is the identification of a "transformation gap" through the rejection of structural invariance. Specifically, the professional group demonstrates a stable ability to convert individual confidence (CSE) into collective capability (CCE), whereas the student group shows a fragmented connection between these two levels.

Furthermore, the research highlights high cognitive volatility and a phenomenon of "premature coupling" within the student group. Students frequently rush into solution generation before goals are fully clarified, indicating that their thinking lacks the logical boundaries characteristic of professionals. Consequently, collected data often remains superficial and fails to be effectively internalized as actionable design knowledge. Based on these evidence-based insights, we advocate for a strategic pedagogical transformation in design education. This transition involves a shift from tool-centric training toward the rigorous cultivation of structured Design Thinking, alongside an explicit focus on developing students’ professional self-identity and psychological resilience. By stabilizing cognitive processes and bridging the identified transformation gap, educators can equip students with the technical methodologies and psychological strategies necessary to achieve a successful transition from academic potential to sustained professional excellence in innovation.

\section{Statements and Declarations}
The authors have no competing interests to declare that are relevant to the content of this article.

\bibliographystyle{unsrt}
\bibliography{references}

@book{amabileCreativityContextUpdate2018,
  title = {Creativity in {{Context}}: {{Update}} to {{The Social Psychology}} of {{Creativity}}},
  shorttitle = {Creativity in {{Context}}},
  author = {Amabile, Teresa M. and Amabile, Teresa M. and Collins, Mary Ann and Conti, Regina and Phillips, Elise and Picariello, Martha and Ruscio, John and Whitney, Dean},
  year = 2018,
  month = may,
  edition = {1},
  publisher = {Routledge},
  doi = {10.4324/9780429501234},
  urldate = {2026-01-19},
  isbn = {978-0-429-50123-4},
  langid = {english}
}

@article{amabileDynamicComponentialModel2016,
  title = {The Dynamic Componential Model of Creativity and Innovation in Organizations: {{Making}} Progress, Making Meaning},
  shorttitle = {The Dynamic Componential Model of Creativity and Innovation in Organizations},
  author = {Amabile, Teresa M. and Pratt, Michael G.},
  year = 2016,
  journal = {Research in Organizational Behavior},
  volume = {36},
  pages = {157--183},
  issn = {01913085},
  doi = {10.1016/j.riob.2016.10.001},
  urldate = {2026-01-20},
  langid = {english}
}

@article{atmanEngineeringDesignProcesses2007,
  title = {Engineering {{Design Processes}}: {{A Comparison}} of {{Students}} and {{Expert Practitioners}}},
  shorttitle = {Engineering {{Design Processes}}},
  author = {Atman, Cynthia J. and Adams, Robin S. and Cardella, Monica E. and Turns, Jennifer and Mosborg, Susan and Saleem, Jason},
  year = 2007,
  month = oct,
  journal = {Journal of Engineering Education},
  volume = {96},
  number = {4},
  pages = {359--379},
  issn = {1069-4730, 2168-9830},
  doi = {10.1002/j.2168-9830.2007.tb00945.x},
  urldate = {2026-01-20},
  copyright = {http://onlinelibrary.wiley.com/termsAndConditions\#vor},
  langid = {english}
}

@article{banduraExerciseHumanAgency2000,
  title = {Exercise of {{Human Agency Through Collective Efficacy}}},
  author = {Bandura, Albert},
  year = 2000,
  month = jun,
  journal = {Current Directions in Psychological Science},
  volume = {9},
  number = {3},
  pages = {75--78},
  issn = {0963-7214, 1467-8721},
  doi = {10.1111/1467-8721.00064},
  urldate = {2026-02-07},
  copyright = {https://journals.sagepub.com/page/policies/text-and-data-mining-license},
  langid = {english}
}

@book{banduraSelfefficacyExerciseControl2012,
  title = {Self-Efficacy: The Exercise of Control},
  shorttitle = {Self-Efficacy},
  author = {Bandura, Albert},
  year = 2012,
  edition = {12. print},
  publisher = {Freeman},
  address = {New York},
  isbn = {978-0-7167-2626-5 978-0-7167-2850-4},
  langid = {english}
}

@article{bareghehMultidisciplinaryDefinitionInnovation2009,
  title = {Towards a Multidisciplinary Definition of Innovation},
  author = {Baregheh, Anahita and Rowley, Jennifer and Sambrook, Sally},
  year = 2009,
  month = sep,
  journal = {Management Decision},
  volume = {47},
  number = {8},
  pages = {1323--1339},
  issn = {0025-1747},
  doi = {10.1108/00251740910984578},
  urldate = {2026-02-03},
  copyright = {https://www.emerald.com/insight/site-policies},
  langid = {english}
}

@article{beckmanInnovationLearningProcess2007,
  title = {Innovation as a {{Learning Process}}: {{Embedding Design Thinking}}},
  shorttitle = {Innovation as a {{Learning Process}}},
  author = {Beckman, Sara L. and Barry, Michael},
  year = 2007,
  month = oct,
  journal = {California Management Review},
  volume = {50},
  number = {1},
  pages = {25--56},
  issn = {0008-1256, 2162-8564},
  doi = {10.2307/41166415},
  urldate = {2026-01-20},
  copyright = {https://journals.sagepub.com/page/policies/text-and-data-mining-license},
  langid = {english}
}

@article{beeftinkBeingSuccessfulCreative2012,
  title = {Being {{Successful}} in a {{Creative Profession}}: {{The Role}} of {{Innovative Cognitive Style}}, {{Self-Regulation}}, and {{Self-Efficacy}}},
  shorttitle = {Being {{Successful}} in a {{Creative Profession}}},
  author = {Beeftink, Flora and Van Eerde, Wendelien and Rutte, Christel G. and Bertrand, J. Will M.},
  year = 2012,
  month = mar,
  journal = {Journal of Business and Psychology},
  volume = {27},
  number = {1},
  pages = {71--81},
  issn = {0889-3268, 1573-353X},
  doi = {10.1007/s10869-011-9214-9},
  urldate = {2025-11-01},
  langid = {english}
}

@book{bollenStructuralEquationsLatent1989a,
  title = {Structural Equations with Latent Variables},
  author = {Bollen, Kenneth A.},
  year = 1989,
  series = {Wiley Series in Probability and Mathematical Statistics. {{Applied}} Probability and Statistics},
  publisher = {Wiley},
  address = {New York},
  doi = {10.1002/9781118619179},
  isbn = {978-0-471-01171-2},
  langid = {english}
}

@article{buchananWickedProblemsDesign1992,
  title = {Wicked {{Problems}} in {{Design Thinking}}},
  author = {Buchanan, Richard},
  year = 1992,
  journal = {Design Issues},
  volume = {8},
  number = {2},
  eprint = {1511637},
  eprinttype = {jstor},
  pages = {5},
  issn = {07479360},
  doi = {10.2307/1511637},
  urldate = {2026-01-20},
  langid = {english}
}

@article{byrneTestingEquivalenceFactor1989,
  title = {Testing for the Equivalence of Factor Covariance and Mean Structures: {{The}} Issue of Partial Measurement Invariance.},
  shorttitle = {Testing for the Equivalence of Factor Covariance and Mean Structures},
  author = {Byrne, Barbara M. and Shavelson, Richard J. and Muth{\'e}n, Bengt},
  year = 1989,
  month = may,
  journal = {Psychological Bulletin},
  volume = {105},
  number = {3},
  pages = {456--466},
  issn = {1939-1455, 0033-2909},
  doi = {10.1037/0033-2909.105.3.456},
  urldate = {2026-01-08},
  langid = {english}
}

@article{caiDevelopmentValidationScale2023,
  title = {The Development and Validation of the Scale of Design Thinking for Teaching ({{SDTT}})},
  author = {Cai, Yuyang and Yang, Yan},
  year = 2023,
  month = jun,
  journal = {Thinking Skills and Creativity},
  volume = {48},
  pages = {101255},
  issn = {18711871},
  doi = {10.1016/j.tsc.2023.101255},
  urldate = {2026-02-11},
  langid = {english}
}

@article{carberryMeasuringEngineeringDesign2010,
  title = {Measuring {{Engineering Design Self}}-{{Efficacy}}},
  author = {Carberry, Adam R. and Lee, Hee-Sun and Ohland, Matthew W.},
  year = 2010,
  month = jan,
  journal = {Journal of Engineering Education},
  volume = {99},
  number = {1},
  pages = {71--79},
  issn = {1069-4730, 2168-9830},
  doi = {10.1002/j.2168-9830.2010.tb01043.x},
  urldate = {2026-02-11},
  copyright = {http://onlinelibrary.wiley.com/termsAndConditions\#vor},
  langid = {english}
}

@article{carlgrenDesignThinkingExploring2014,
  title = {Design {{Thinking}}: {{Exploring Values}} and {{Effects}} from an {{Innovation Capability Perspective}}},
  shorttitle = {Design {{Thinking}}},
  author = {Carlgren, Lisa and Elmquist, Maria and Rauth, Ingo},
  year = 2014,
  month = sep,
  journal = {The Design Journal},
  volume = {17},
  number = {3},
  pages = {403--423},
  issn = {1460-6925, 1756-3062},
  doi = {10.2752/175630614X13982745783000},
  urldate = {2026-01-20},
  langid = {english}
}

@article{carsonReliabilityValidityFactor2005,
  title = {Reliability, {{Validity}}, and {{Factor Structure}} of the {{Creative Achievement Questionnaire}}},
  author = {Carson, Shelley H. and Peterson, Jordan B. and Higgins, Daniel M.},
  year = 2005,
  month = feb,
  journal = {Creativity Research Journal},
  volume = {17},
  number = {1},
  pages = {37--50},
  issn = {1040-0419, 1532-6934},
  doi = {10.1207/s15326934crj1701_4},
  urldate = {2025-11-01},
  langid = {english}
}

@article{chengAntecedentsCollectiveCreative2014,
  title = {The Antecedents of Collective Creative Efficacy for Information System Development Teams},
  author = {Cheng, Hsiu-Hua and Yang, Heng-Li},
  year = 2014,
  month = jul,
  journal = {Journal of Engineering and Technology Management},
  volume = {33},
  pages = {1--17},
  issn = {09234748},
  doi = {10.1016/j.jengtecman.2013.12.001},
  urldate = {2026-02-03},
  langid = {english}
}

@article{chenSystemsTheoryMotivated2006,
  title = {Toward a {{Systems Theory}} of {{Motivated Behavior}} in {{Work Teams}}},
  author = {Chen, Gilad and Kanfer, Ruth},
  year = 2006,
  month = jan,
  journal = {Research in Organizational Behavior},
  volume = {27},
  pages = {223--267},
  issn = {01913085},
  doi = {10.1016/S0191-3085(06)27006-0},
  urldate = {2026-01-11},
  copyright = {https://www.elsevier.com/tdm/userlicense/1.0/},
  langid = {english}
}

@article{cheungEvaluatingGoodnessofFitIndexes2002a,
  title = {Evaluating {{Goodness-of-Fit Indexes}} for {{Testing Measurement Invariance}}},
  author = {Cheung, Gordon W. and Rensvold, Roger B.},
  year = 2002,
  month = apr,
  journal = {Structural Equation Modeling: A Multidisciplinary Journal},
  volume = {9},
  number = {2},
  pages = {233--255},
  issn = {1070-5511, 1532-8007},
  doi = {10.1207/S15328007SEM0902_5},
  urldate = {2026-01-08},
  langid = {english}
}

@article{christiaansCreativityDesignEngineering2005,
  title = {Creativity in {{Design Engineering}} and the {{Role}} of {{Knowledge}}: {{Modelling}} the {{Expert}}},
  shorttitle = {Creativity in {{Design Engineering}} and the {{Role}} of {{Knowledge}}},
  author = {Christiaans, Henri and Venselaar, Kees},
  year = 2005,
  month = jan,
  journal = {International Journal of Technology and Design Education},
  volume = {15},
  number = {3},
  pages = {217--236},
  issn = {0957-7572, 1573-1804},
  doi = {10.1007/s10798-004-1904-4},
  urldate = {2026-01-20},
  copyright = {http://www.springer.com/tdm},
  langid = {english}
}

@article{crossDesignerlyWaysKnowing2001,
  title = {Designerly {{Ways}} of {{Knowing}}: {{Design Discipline Versus Design Science}}},
  shorttitle = {Designerly {{Ways}} of {{Knowing}}},
  author = {Cross, Nigel},
  year = 2001,
  month = jul,
  journal = {Design Issues},
  volume = {17},
  number = {3},
  pages = {49--55},
  issn = {0747-9360, 1531-4790},
  doi = {10.1162/074793601750357196},
  urldate = {2026-01-10},
  langid = {english}
}

@book{crossDesignThinkingUnderstanding2016,
  title = {Design Thinking: Understanding How Designers Think and Work},
  shorttitle = {Design Thinking},
  author = {Cross, Nigel},
  year = 2016,
  edition = {Reprinted},
  publisher = {Bloomsbury Academic},
  address = {London Oxford},
  isbn = {978-1-84788-636-1 978-1-84788-637-8},
  langid = {english}
}

@article{crossExpertiseDesignOverview2004,
  title = {Expertise in Design: An Overview},
  shorttitle = {Expertise in Design},
  author = {Cross, Nigel},
  year = 2004,
  month = sep,
  journal = {Design Studies},
  volume = {25},
  number = {5},
  pages = {427--441},
  issn = {0142694X},
  doi = {10.1016/j.destud.2004.06.002},
  urldate = {2026-01-11},
  copyright = {https://www.elsevier.com/tdm/userlicense/1.0/},
  langid = {english}
}

@article{dollConfirmatoryFactorAnalysis1994,
  title = {A {{Confirmatory Factor Analysis}} of the {{End-User Computing Satisfaction Instrument}}},
  author = {Doll, William J. and Xia, Weidong and Torkzadeh, Gholamreza},
  year = 1994,
  month = dec,
  journal = {MIS Quarterly},
  volume = {18},
  number = {4},
  pages = {453--461},
  issn = {0276-7783, 2162-9730},
  doi = {10.2307/249524},
  urldate = {2026-01-08},
  langid = {english}
}

@article{dorstCoreDesignThinking2011,
  title = {The Core of `Design Thinking' and Its Application},
  author = {Dorst, Kees},
  year = 2011,
  month = nov,
  journal = {Design Studies},
  volume = {32},
  number = {6},
  pages = {521--532},
  issn = {0142694X},
  doi = {10.1016/j.destud.2011.07.006},
  urldate = {2026-01-10},
  copyright = {https://www.elsevier.com/tdm/userlicense/1.0/},
  langid = {english}
}

@article{dorstCreativityDesignProcess2001,
  title = {Creativity in the Design Process: Co-Evolution of Problem--Solution},
  shorttitle = {Creativity in the Design Process},
  author = {Dorst, Kees and Cross, Nigel},
  year = 2001,
  month = sep,
  journal = {Design Studies},
  volume = {22},
  number = {5},
  pages = {425--437},
  issn = {0142694X},
  doi = {10.1016/S0142-694X(01)00009-6},
  urldate = {2026-01-10},
  copyright = {https://www.elsevier.com/tdm/userlicense/1.0/},
  langid = {english}
}

@book{dreyfusMindMachinePower1988,
  title = {Mind over Machine: The Power of Human Intuition and Expertise in the Era of the Computer},
  shorttitle = {Mind over Machine},
  author = {Dreyfus, Hubert L. and Dreyfus, Stuart E.},
  year = 1988,
  edition = {1. paperback ed},
  publisher = {The Free Pr},
  address = {New York},
  isbn = {978-0-02-908061-0},
  langid = {english}
}

@article{dunningFlawedSelfAssessmentImplications2004,
  title = {Flawed {{Self-Assessment}}: {{Implications}} for {{Health}}, {{Education}}, and the {{Workplace}}},
  shorttitle = {Flawed {{Self-Assessment}}},
  author = {Dunning, David and Heath, Chip and Suls, Jerry M.},
  year = 2004,
  month = dec,
  journal = {Psychological Science in the Public Interest},
  volume = {5},
  number = {3},
  pages = {69--106},
  issn = {1529-1006, 1539-6053},
  doi = {10.1111/j.1529-1006.2004.00018.x},
  urldate = {2026-01-10},
  copyright = {https://journals.sagepub.com/page/policies/text-and-data-mining-license},
  langid = {english}
}

@article{dymEngineeringDesignThinking2005,
  title = {Engineering {{Design Thinking}}, {{Teaching}}, and {{Learning}}},
  author = {Dym, Clive L. and Agogino, Alice M. and Eris, Ozgur and Frey, Daniel D. and Leifer, Larry J.},
  year = 2005,
  month = jan,
  journal = {Journal of Engineering Education},
  volume = {94},
  number = {1},
  pages = {103--120},
  issn = {10694730},
  doi = {10.1002/j.2168-9830.2005.tb00832.x},
  urldate = {2026-01-20},
  copyright = {http://doi.wiley.com/10.1002/tdm\_license\_1.1},
  langid = {english}
}

@article{elsbachDesignThinkingOrganizational2018,
  title = {Design {{Thinking}} and {{Organizational Culture}}: {{A Review}} and {{Framework}} for {{Future Research}}},
  shorttitle = {Design {{Thinking}} and {{Organizational Culture}}},
  author = {Elsbach, Kimberly D. and Stigliani, Ileana},
  year = 2018,
  month = jul,
  journal = {Journal of Management},
  volume = {44},
  number = {6},
  pages = {2274--2306},
  issn = {0149-2063, 1557-1211},
  doi = {10.1177/0149206317744252},
  urldate = {2026-01-20},
  langid = {english}
}

@book{ericssonCambridgeHandbookExpertise2013,
  title = {The {{Cambridge}} Handbook of Expertise and Expert Performance},
  editor = {Ericsson, Karl Anders and Charness, Neil H. and Feltovich, Paul J. and Hoffman, Robert R.},
  year = 2013,
  edition = {11. printing},
  publisher = {Cambridge University Press},
  address = {Cambridge},
  isbn = {978-0-521-60081-1 978-0-521-84097-2},
  langid = {english}
}

@article{fornellEvaluatingStructuralEquation1981,
  title = {Evaluating {{Structural Equation Models}} with {{Unobservable Variables}} and {{Measurement Error}}},
  author = {Fornell, Claes and Larcker, David F.},
  year = 1981,
  month = feb,
  journal = {Journal of Marketing Research},
  volume = {18},
  number = {1},
  pages = {39--50},
  issn = {0022-2437, 1547-7193},
  doi = {10.1177/002224378101800104},
  urldate = {2024-07-21},
  copyright = {http://journals.sagepub.com/page/policies/text-and-data-mining-license},
  langid = {english}
}

@article{gongEmployeeLearningOrientation2009,
  title = {Employee {{Learning Orientation}}, {{Transformational Leadership}}, and {{Employee Creativity}}: {{The Mediating Role}} of {{Employee Creative Self-Efficacy}}},
  shorttitle = {Employee {{Learning Orientation}}, {{Transformational Leadership}}, and {{Employee Creativity}}},
  author = {Gong, Yaping and Huang, Jia-Chi and Farh, Jiing-Lih},
  year = 2009,
  month = aug,
  journal = {Academy of Management Journal},
  volume = {52},
  number = {4},
  pages = {765--778},
  issn = {0001-4273, 1948-0989},
  doi = {10.5465/amj.2009.43670890},
  urldate = {2026-01-20},
  langid = {english}
}

@book{hairMultivariateDataAnalysis2010,
  title = {Multivariate Data Analysis},
  editor = {Hair, Joseph F. and Black, William C. and Babin, Barry J. and Anderson, Rolph E.},
  year = 2010,
  edition = {7. ed},
  publisher = {Pearson Prentice Hall},
  address = {Upper Saddle River, NJ},
  isbn = {978-0-13-813263-7},
  langid = {english}
}

@book{hayesIntroductionMediationModeration2018,
  title = {Introduction to Mediation, Moderation, and Conditional Process Analysis: A Regression-Based Approach},
  shorttitle = {Introduction to Mediation, Moderation, and Conditional Process Analysis},
  author = {Hayes, Andrew F.},
  year = 2018,
  series = {Methodology in the Social Sciences},
  edition = {Second edition},
  publisher = {Guilford Press},
  address = {New York},
  isbn = {978-1-4625-3467-8},
  langid = {english}
}

@article{hinkinReviewScaleDevelopment1995,
  title = {A {{Review}} of {{Scale Development Practices}} in the {{Study}} of {{Organizations}}},
  author = {Hinkin, Timothy R.},
  year = 1995,
  month = oct,
  journal = {Journal of Management},
  volume = {21},
  number = {5},
  pages = {967--988},
  issn = {0149-2063, 1557-1211},
  doi = {10.1177/014920639502100509},
  urldate = {2026-01-21},
  copyright = {https://journals.sagepub.com/page/policies/text-and-data-mining-license},
  langid = {english}
}

@article{huCutoffCriteriaFit1999,
  title = {Cutoff Criteria for Fit Indexes in Covariance Structure Analysis: {{Conventional}} Criteria versus New Alternatives},
  shorttitle = {Cutoff Criteria for Fit Indexes in Covariance Structure Analysis},
  author = {Hu, Li-tze and Bentler, Peter M.},
  year = 1999,
  month = jan,
  journal = {Structural Equation Modeling: A Multidisciplinary Journal},
  volume = {6},
  number = {1},
  pages = {1--55},
  issn = {1070-5511, 1532-8007},
  doi = {10.1080/10705519909540118},
  urldate = {2026-01-08},
  langid = {english}
}

@article{jablokowInvestigatingInfluenceDesigners2019,
  title = {Investigating the {{Influence}} of {{Designers}}' {{Cognitive Characteristics}} and {{Interaction Behaviors}} in {{Design Concept Generation}}},
  author = {Jablokow, Kathryn W. and Sonalkar, Neeraj and Edelman, Jonathan and Mabogunje, Ade and Leifer, Larry},
  year = 2019,
  month = sep,
  journal = {Journal of Mechanical Design},
  volume = {141},
  number = {9},
  pages = {091101},
  issn = {1050-0472, 1528-9001},
  doi = {10.1115/1.4043316},
  urldate = {2026-01-19},
  langid = {english}
}

@article{johansson-skoldbergDesignThinkingPresent2013,
  title = {Design {{Thinking}}: {{Past}}, {{Present}} and {{Possible Futures}}},
  shorttitle = {Design {{Thinking}}},
  author = {Johansson-Sk{\"o}ldberg, Ulla and Woodilla, Jill and {\c C}etinkaya, Mehves},
  year = 2013,
  month = jun,
  journal = {Creativity and Innovation Management},
  volume = {22},
  number = {2},
  pages = {121--146},
  issn = {0963-1690, 1467-8691},
  doi = {10.1111/caim.12023},
  urldate = {2026-01-10},
  copyright = {http://onlinelibrary.wiley.com/termsAndConditions\#vor},
  langid = {english}
}

@article{joreskogSimultaneousFactorAnalysis1971,
  title = {Simultaneous {{Factor Analysis}} in {{Several Populations}}},
  author = {J{\"o}reskog, K. G.},
  year = 1971,
  month = dec,
  journal = {Psychometrika},
  volume = {36},
  number = {4},
  pages = {409--426},
  issn = {0033-3123, 1860-0980},
  doi = {10.1007/BF02291366},
  urldate = {2026-01-08},
  copyright = {https://www.cambridge.org/core/terms},
  langid = {english}
}

@article{kleinsmannBarriersEnablersCreating2008,
  title = {Barriers and Enablers for Creating Shared Understanding in Co-Design Projects},
  author = {Kleinsmann, Maaike and Valkenburg, Rianne},
  year = 2008,
  month = jul,
  journal = {Design Studies},
  volume = {29},
  number = {4},
  pages = {369--386},
  issn = {0142694X},
  doi = {10.1016/j.destud.2008.03.003},
  urldate = {2026-01-20},
  copyright = {https://www.elsevier.com/tdm/userlicense/1.0/},
  langid = {english}
}

@book{klinePrinciplesPracticeStructural2016,
  title = {Principles and Practice of Structural Equation Modeling},
  author = {Kline, Rex B.},
  year = 2016,
  series = {Methodology in the Social Sciences},
  edition = {Fourth edition},
  publisher = {The Guilford Press},
  address = {New York London},
  isbn = {978-1-4625-2335-1 978-1-4625-2336-8},
  langid = {english}
}

@article{kreitlerSelfPerceivedCreativityPerspective2009a,
  title = {Self-{{Perceived Creativity}}: {{The Perspective}} of {{Design}}},
  shorttitle = {Self-{{Perceived Creativity}}},
  author = {Kreitler, Shulamith and Casakin, Hernan},
  year = 2009,
  month = jan,
  journal = {European Journal of Psychological Assessment},
  volume = {25},
  number = {3},
  pages = {194--203},
  issn = {1015-5759, 2151-2426},
  doi = {10.1027/1015-5759.25.3.194},
  urldate = {2026-02-11},
  langid = {english}
}

@article{krugerSolutionDrivenProblem2006,
  title = {Solution Driven versus Problem Driven Design: Strategies and Outcomes},
  shorttitle = {Solution Driven versus Problem Driven Design},
  author = {Kruger, Corinne and Cross, Nigel},
  year = 2006,
  month = sep,
  journal = {Design Studies},
  volume = {27},
  number = {5},
  pages = {527--548},
  issn = {0142694X},
  doi = {10.1016/j.destud.2006.01.001},
  urldate = {2025-11-01},
  copyright = {https://www.elsevier.com/tdm/userlicense/1.0/},
  langid = {english}
}

@article{lambMeasuringCreativeSelfefficacy2025,
  title = {Measuring Creative Self-Efficacy: {{Instrument}} Development and Validation},
  shorttitle = {Measuring Creative Self-Efficacy},
  author = {Lamb, Kristen N. and Boedeker, Peter and Kettler, Todd},
  year = 2025,
  month = jun,
  journal = {Thinking Skills and Creativity},
  volume = {56},
  pages = {101738},
  issn = {18711871},
  doi = {10.1016/j.tsc.2024.101738},
  urldate = {2025-11-01},
  langid = {english}
}

@article{liedtkaPerspectiveLinkingDesign2015,
  title = {Perspective: {{Linking Design Thinking}} with {{Innovation Outcomes}} through {{Cognitive Bias Reduction}}},
  shorttitle = {Perspective},
  author = {Liedtka, Jeanne},
  year = 2015,
  month = nov,
  journal = {Journal of Product Innovation Management},
  volume = {32},
  number = {6},
  pages = {925--938},
  issn = {0737-6782, 1540-5885},
  doi = {10.1111/jpim.12163},
  urldate = {2026-01-21},
  copyright = {http://onlinelibrary.wiley.com/termsAndConditions\#vor},
  langid = {english}
}

@article{luRelationshipStudentDesign2015,
  title = {The Relationship between Student Design Cognition Types and Creative Design Outcomes},
  author = {Lu, Chia-Chen},
  year = 2015,
  month = jan,
  journal = {Design Studies},
  volume = {36},
  pages = {59--76},
  issn = {0142694X},
  doi = {10.1016/j.destud.2014.08.002},
  urldate = {2025-10-31},
  langid = {english}
}

@article{mackinnonConfidenceLimitsIndirect2004,
  title = {Confidence {{Limits}} for the {{Indirect Effect}}: {{Distribution}} of the {{Product}} and {{Resampling Methods}}},
  shorttitle = {Confidence {{Limits}} for the {{Indirect Effect}}},
  author = {MacKinnon, David P. and Lockwood, Chondra M. and Williams, Jason},
  year = 2004,
  month = jan,
  journal = {Multivariate Behavioral Research},
  volume = {39},
  number = {1},
  pages = {99--128},
  issn = {0027-3171, 1532-7906},
  doi = {10.1207/s15327906mbr3901_4},
  urldate = {2026-01-08},
  langid = {english}
}

@article{marshSearchGoldenRules2004,
  title = {In {{Search}} of {{Golden Rules}}: {{Comment}} on {{Hypothesis-Testing Approaches}} to {{Setting Cutoff Values}} for {{Fit Indexes}} and {{Dangers}} in {{Overgeneralizing Hu}} and {{Bentler}}'s (1999) {{Findings}}},
  shorttitle = {In {{Search}} of {{Golden Rules}}},
  author = {Marsh, Herbert W. and Hau, Kit-Tai and Wen, Zhonglin},
  year = 2004,
  month = jul,
  journal = {Structural Equation Modeling: A Multidisciplinary Journal},
  volume = {11},
  number = {3},
  pages = {320--341},
  issn = {1070-5511, 1532-8007},
  doi = {10.1207/s15328007sem1103_2},
  urldate = {2026-01-08},
  langid = {english}
}

@book{martinDesignBusinessWhy2009,
  title = {Design of {{Business}}: {{Why Design Thinking}} Is the {{Next Competitive Advantage}}},
  shorttitle = {Design of {{Business}}},
  author = {Martin, Roger L.},
  year = 2009,
  publisher = {Harvard Business Review Press},
  address = {Boston},
  isbn = {978-1-4221-7780-8 978-1-4221-5511-0},
  langid = {english}
}

@article{mathisenCreativeSelfefficacyIntervention2009,
  title = {Creative Self-Efficacy: {{An}} Intervention Study},
  shorttitle = {Creative Self-Efficacy},
  author = {Mathisen, Gro Ellen and Bronnick, Kolbjorn S.},
  year = 2009,
  month = jan,
  journal = {International Journal of Educational Research},
  volume = {48},
  number = {1},
  pages = {21--29},
  issn = {08830355},
  doi = {10.1016/j.ijer.2009.02.009},
  urldate = {2026-02-03},
  copyright = {https://www.elsevier.com/tdm/userlicense/1.0/},
  langid = {english}
}

@article{mcnallyProductInnovativenessDimensions2010,
  title = {Product {{Innovativeness Dimensions}} and {{Their Relationships}} with {{Product Advantage}}, {{Product Financial Performance}}, and {{Project Protocol}}},
  author = {McNally, Regina C. and Cavusgil, Erin and Calantone, Roger J.},
  year = 2010,
  month = dec,
  journal = {Journal of Product Innovation Management},
  volume = {27},
  number = {7},
  pages = {991--1006},
  issn = {0737-6782, 1540-5885},
  doi = {10.1111/j.1540-5885.2010.00766.x},
  urldate = {2026-02-11},
  copyright = {http://onlinelibrary.wiley.com/termsAndConditions\#vor},
  langid = {english}
}

@article{oxmanEducatingDesignerlyThinker1999,
  title = {Educating the Designerly Thinker},
  author = {Oxman, Rivka},
  year = 1999,
  month = mar,
  journal = {Design Studies},
  volume = {20},
  number = {2},
  pages = {105--122},
  issn = {0142694X},
  doi = {10.1016/S0142-694X(98)00029-5},
  urldate = {2026-01-20},
  copyright = {https://www.elsevier.com/tdm/userlicense/1.0/},
  langid = {english}
}

@article{podsakoffCommonMethodBiases2003,
  title = {Common Method Biases in Behavioral Research: {{A}} Critical Review of the Literature and Recommended Remedies.},
  shorttitle = {Common Method Biases in Behavioral Research},
  author = {Podsakoff, Philip M. and MacKenzie, Scott B. and Lee, Jeong-Yeon and Podsakoff, Nathan P.},
  year = 2003,
  journal = {Journal of Applied Psychology},
  volume = {88},
  number = {5},
  pages = {879--903},
  issn = {1939-1854, 0021-9010},
  doi = {10.1037/0021-9010.88.5.879},
  urldate = {2026-01-08},
  langid = {english}
}

@article{podsakoffSourcesMethodBias2012,
  title = {Sources of {{Method Bias}} in {{Social Science Research}} and {{Recommendations}} on {{How}} to {{Control It}}},
  author = {Podsakoff, Philip M. and MacKenzie, Scott B. and Podsakoff, Nathan P.},
  year = 2012,
  month = jan,
  journal = {Annual Review of Psychology},
  volume = {63},
  number = {1},
  pages = {539--569},
  issn = {0066-4308, 1545-2085},
  doi = {10.1146/annurev-psych-120710-100452},
  urldate = {2026-01-08},
  langid = {english}
}

@article{preacherAsymptoticResamplingStrategies2008a,
  title = {Asymptotic and Resampling Strategies for Assessing and Comparing Indirect Effects in Multiple Mediator Models},
  author = {Preacher, Kristopher J. and Hayes, Andrew F.},
  year = 2008,
  month = aug,
  journal = {Behavior Research Methods},
  volume = {40},
  number = {3},
  pages = {879--891},
  issn = {1554-351X, 1554-3528},
  doi = {10.3758/BRM.40.3.879},
  urldate = {2026-01-08},
  copyright = {http://www.springer.com/tdm},
  langid = {english}
}

@book{schonReflectivePractitionerHow1983,
  title = {The Reflective Practitioner: How Professionals Think in Action},
  shorttitle = {The Reflective Practitioner},
  author = {Sch{\"o}n, Donald A.},
  year = 1983,
  publisher = {Basic Books},
  address = {New York},
  isbn = {978-0-465-06874-6 978-0-465-06878-4},
  langid = {english}
}

@article{shinWhenEducationalSpecialization2007,
  title = {When Is Educational Specialization Heterogeneity Related to Creativity in Research and Development Teams? {{Transformational}} Leadership as a Moderator.},
  shorttitle = {When Is Educational Specialization Heterogeneity Related to Creativity in Research and Development Teams?},
  author = {Shin, Shung J. and Zhou, Jing},
  year = 2007,
  month = nov,
  journal = {Journal of Applied Psychology},
  volume = {92},
  number = {6},
  pages = {1709--1721},
  issn = {1939-1854, 0021-9010},
  doi = {10.1037/0021-9010.92.6.1709},
  urldate = {2026-01-11},
  langid = {english}
}

@book{simonSciencesArtificial2019,
  title = {The {{Sciences}} of the {{Artificial}}},
  author = {Simon, Herbert A.},
  year = 2019,
  month = aug,
  publisher = {The MIT Press},
  doi = {10.7551/mitpress/12107.001.0001},
  urldate = {2026-01-19},
  isbn = {978-0-262-35474-5},
  langid = {english}
}

@article{stajkovicCollectiveEfficacyGroup2009,
  title = {Collective Efficacy, Group Potency, and Group Performance: {{Meta-analyses}} of Their Relationships, and Test of a Mediation Model.},
  shorttitle = {Collective Efficacy, Group Potency, and Group Performance},
  author = {Stajkovic, Alexander D. and Lee, Dongseop and Nyberg, Anthony J.},
  year = 2009,
  journal = {Journal of Applied Psychology},
  volume = {94},
  number = {3},
  pages = {814--828},
  issn = {1939-1854, 0021-9010},
  doi = {10.1037/a0015659},
  urldate = {2026-01-11},
  langid = {english}
}

@article{steenkampAssessingMeasurementInvariance1998,
  title = {Assessing {{Measurement Invariance}} in {{Cross}}-{{National Consumer Research}}},
  author = {Steenkamp, Jan-Benedict~E.~M. and Baumgartner, Hans},
  year = 1998,
  month = jun,
  journal = {Journal of Consumer Research},
  volume = {25},
  number = {1},
  pages = {78--107},
  issn = {0093-5301, 1537-5277},
  doi = {10.1086/209528},
  urldate = {2026-01-08},
  langid = {english}
}

@article{tierneyCreativeSelfefficacyDevelopment2011,
  title = {Creative Self-Efficacy Development and Creative Performance over Time.},
  author = {Tierney, Pamela and Farmer, Steven M.},
  year = 2011,
  journal = {Journal of Applied Psychology},
  volume = {96},
  number = {2},
  pages = {277--293},
  issn = {1939-1854, 0021-9010},
  doi = {10.1037/a0020952},
  urldate = {2026-01-21},
  langid = {english}
}

@article{tierneyCreativeSelfEfficacyIts2002,
  title = {Creative {{Self-Efficacy}}: {{Its Potential Antecedents}} and {{Relationship}} to {{Creative Performance}}},
  shorttitle = {{{CREATIVE SELF-EFFICACY}}},
  author = {Tierney, P. and Farmer, S. M.},
  year = 2002,
  month = dec,
  journal = {Academy of Management Journal},
  volume = {45},
  number = {6},
  pages = {1137--1148},
  issn = {0001-4273, 1948-0989},
  doi = {10.2307/3069429},
  urldate = {2025-11-03},
  langid = {english}
}

@article{vignoliDesignThinkingMindset2023,
  title = {Design Thinking Mindset: Scale Development and Validation},
  shorttitle = {Design Thinking Mindset},
  author = {Vignoli, Matteo and Dosi, Clio and Balboni, Bernardo},
  year = 2023,
  month = jun,
  journal = {Studies in Higher Education},
  volume = {48},
  number = {6},
  pages = {926--940},
  issn = {0307-5079, 1470-174X},
  doi = {10.1080/03075079.2023.2172566},
  urldate = {2026-02-11},
  langid = {english}
}

@article{zouCognitiveCharacteristicsInnovation2023a,
  title = {Cognitive {{Characteristics}} of an {{Innovation Team}} and {{Collaborative Innovation Performance}}: {{The Mediating Role}} of {{Cooperative Behavior}} and the {{Moderating Role}} of {{Team Innovation Efficacy}}},
  shorttitle = {Cognitive {{Characteristics}} of an {{Innovation Team}} and {{Collaborative Innovation Performance}}},
  author = {Zou, Mi and Liu, Peng and Wu, Xuan and Zhou, Wei and Jin, Yuan and Xu, Meiqi},
  year = 2023,
  month = jul,
  journal = {Sustainability},
  volume = {15},
  number = {14},
  pages = {10951},
  issn = {2071-1050},
  doi = {10.3390/su151410951},
  urldate = {2026-02-11},
  langid = {english}
}

\newpage
\appendix
\section*{Appendix}
\setcounter{table}{0}
\renewcommand{\thetable}{\Alph{section}\arabic{table}}

\begin{table}[h] 
\centering
\vspace*{\fill} 
\caption{Questionnaire Based on Design Cognition Strategies, Design Thinking, Creative Efficacy and Design Creativity \& Innovation}
\label{tab: Questionnaire}
\begin{tabularx}{\linewidth}{@{} l l L @{}} 
\toprule
Variables & Items & Content \\ 
\midrule
\multirow{5}{*}{Problem-driven Design} & PD1 & I list design-resolvable issues for design goals and direction. \\
& PD2 & I thoroughly formulate my design goals and direction. \\
& PD3 & My ideas result from the design goal and direction that I establish. \\
& PD4 & I evaluate ideas based on the design goal and direction that I establish. \\
& PD5 & I focus on the design goal and direction during the overall design process. \\
\midrule
\multirow{5}{*}{Information-driven Design} & ID1 & I ask questions if the problems in the design task are unclear. \\
& ID2 & I spend a substantial amount of time gathering information. \\
& ID3 & I consider how to gather and organize information. \\
& ID4 & I identify Design-Solvable Issues from massive information. \\
& ID5 & I set design goals and direction from tasks and gathered information. \\
\midrule
\multirow{5}{*}{Solution-driven Design} & SD1 & I primarily gather information to reinforce the idea that I develop. \\
& SD2 & I can consider a number of feasible design ideas simultaneously. \\
& SD3 & I always generate more design ideas than others do. \\
& SD4 & I can generate design ideas in a short amount of time. \\
& SD5 & I can generate various design ideas to provide diversity. \\
\midrule
\multirow{5}{*}{Knowledge-driven Design} & KD1 & I recall similar design problems when reading the task. \\
& KD2 & I use design knowledge to examine design-resolvable problems. \\
& KD3 & I believe that my design knowledge assists me in developing design ideas. \\
& KD4 & I develop design ideas based on similar designs I remember. \\
& KD5 & I depend on prior design experience to generate design ideas. \\
\midrule
\multirow{5}{*}{Design thinking} & DT1 & I optimize and adjust the design through iterative processes and validation. \\
& DT2 & I focus on user experience and improve the design based on feedback. \\
& DT3 & I flexibly apply divergent and convergent thinking based on design needs. \\
& DT4 & I can integrate interdisciplinary knowledge into the design process. \\
& DT5 & I am able to apply design methods in a systematic way to find solutions. \\
\midrule
\multirow{5}{*}{Creative Self-Efficacy} & CSE1 & I have confidence in my ability to solve problems creatively. \\
& CSE2 & I have confidence in my ability to come up with new ideas. \\
& CSE3 & I have confidence in my ability to be a creative person. \\
& CSE4 & I am confident I can generate diverse design concepts for a problem. \\
& CSE5 & I am confident I can turn a concept into a workable prototype or solution. \\
\midrule
\multirow{5}{*}{Collective Creative Efficacy} & CCE1 & I have confidence in the ability of my team to solve problems creatively. \\
& CCE2 & I have confidence in my team's ability to produce new ideas. \\
& CCE3 & As a team, we can achieve most of the creative goals we set for ourselves. \\
& CCE4 & My team is capable of turning new ideas into workable design solutions. \\
& CCE5 & Team members support each other in creative risks and experiments. \\
\midrule
\multirow{5}{*}{Design Creativity \& Innovation} & DCI1 & I am usually able to provide completely new perspectives or solutions. \\
& DCI2 & My designs solve practical problems and meet user needs to some extent. \\
& DCI3 & I consider different design options and solutions. \\
& DCI4 & I optimize my designs solutions based on the collected information. \\
& DCI5 & I re-examine the problem from different perspectives to find new solutions. \\
\bottomrule
\end{tabularx}
\vspace*{\fill} 
\end{table}

\begin{table}
\caption{Measurement Model Results (Factor Loadings, Reliability, and Validity)}
\label{tab: measurement_model}
\begin{tabular*}{\linewidth}{@{\extracolsep{\fill}} l l c c c c c c c c @{}} 
\toprule
Variables & Items & Estimate & S.E. & t-value & Sig.   & Std. & SMC & CR & AVE \\
\midrule
\multirow{5}{*}{Problem-driven Design} 
  & PD1 & 1.000 &       &        &     & 0.678 & 0.460 & \multirow{5}{*}{0.809} & \multirow{5}{*}{0.460} \\
  & PD2 & 1.099 & 0.086 & 12.790 & *** & 0.692 & 0.479 &  &  \\
  & PD3 & 1.012 & 0.083 & 12.162 & *** & 0.652 & 0.425 &  &  \\
  & PD4 & 0.895 & 0.081 & 11.061 & *** & 0.584 & 0.341 &  &  \\
  & PD5 & 1.212 & 0.087 & 13.890 & *** & 0.771 & 0.594 &  &  \\
\midrule
\multirow{5}{*}{Information-driven Design} 
  & ID1 & 1.000 &       &        &     & 0.732 & 0.536 & \multirow{5}{*}{0.838} & \multirow{5}{*}{0.511} \\
  & ID2 & 0.757 & 0.062 & 12.261 & *** & 0.604 & 0.365 &  &  \\
  & ID3 & 1.108 & 0.069 & 16.049 & *** & 0.799 & 0.638 &  &  \\
  & ID4 & 0.850 & 0.067 & 12.595 & *** & 0.620 & 0.384 &  &  \\
  & ID5 & 1.081 & 0.068 & 15.968 & *** & 0.795 & 0.632 &  &  \\
\midrule
\multirow{5}{*}{Solution-driven Design} 
  & SD1 & 1.000 &       &        &     & 0.719 & 0.517 & \multirow{5}{*}{0.874} & \multirow{5}{*}{0.581} \\
  & SD2 & 1.319 & 0.082 & 16.186 & *** & 0.793 & 0.629 &  &  \\
  & SD3 & 1.275 & 0.079 & 16.178 & *** & 0.792 & 0.627 &  &  \\
  & SD4 & 1.243 & 0.076 & 16.294 & *** & 0.798 & 0.637 &  &  \\
  & SD5 & 1.135 & 0.079 & 14.416 & *** & 0.703 & 0.494 &  &  \\
\midrule
\multirow{5}{*}{Knowledge-driven Design} 
  & KD1 & 1.000 &       &        &     & 0.751 & 0.564 & \multirow{5}{*}{0.816} & \multirow{5}{*}{0.472} \\
  & KD2 & 0.891 & 0.066 & 13.405 & *** & 0.663 & 0.440 &  &  \\
  & KD3 & 0.925 & 0.061 & 15.046 & *** & 0.748 & 0.560 &  &  \\
  & KD4 & 0.750 & 0.059 & 12.644 & *** & 0.625 & 0.391 &  &  \\
  & KD5 & 0.809 & 0.063 & 12.885 & *** & 0.637 & 0.406 &  &  \\
\midrule
\multirow{5}{*}{Design thinking} 
  & DT1 & 1.000 &       &        &     & 0.803 & 0.645 & \multirow{5}{*}{0.870} & \multirow{5}{*}{0.576} \\
  & DT2 & 0.999 & 0.053 & 18.808 & *** & 0.794 & 0.630 &  &  \\
  & DT3 & 1.028 & 0.051 & 20.021 & *** & 0.835 & 0.697 &  &  \\
  & DT4 & 0.634 & 0.046 & 13.870 & *** & 0.619 & 0.383 &  &  \\
  & DT5 & 0.788 & 0.047 & 16.706 & *** & 0.723 & 0.523 &  &  \\
\midrule
\multirow{5}{*}{Creative Self-Efficacy} 
  & CSE1 & 1.000 &       &        &     & 0.835 & 0.697 & \multirow{5}{*}{0.869} & \multirow{5}{*}{0.572} \\
  & CSE2 & 0.762 & 0.049 & 15.577 & *** & 0.674 & 0.454 &  &  \\
  & CSE3 & 1.038 & 0.053 & 19.463 & *** & 0.802 & 0.643 &  &  \\
  & CSE4 & 0.953 & 0.052 & 18.470 & *** & 0.770 & 0.593 &  &  \\
  & CSE5 & 0.806 & 0.051 & 15.953 & *** & 0.687 & 0.472 &  &  \\
\midrule
\multirow{5}{*}{Collective Creative Efficacy} 
  & CCE1 & 1.000 &       &        &     & 0.812 & 0.659 & \multirow{5}{*}{0.907} & \multirow{5}{*}{0.662} \\
  & CCE2 & 1.153 & 0.052 & 22.106 & *** & 0.870 & 0.757 &  &  \\
  & CCE3 & 1.048 & 0.048 & 21.895 & *** & 0.864 & 0.746 &  &  \\
  & CCE4 & 1.018 & 0.052 & 19.726 & *** & 0.801 & 0.642 &  &  \\
  & CCE5 & 0.861 & 0.051 & 16.804 & *** & 0.710 & 0.504 &  &  \\
\midrule
\multirow{5}{*}{Design Creativity \& Innovation} 
  & DCI1 & 1.000 &       &        &     & 0.773 & 0.598 & \multirow{5}{*}{0.893} & \multirow{5}{*}{0.626} \\
  & DCI2 & 1.087 & 0.061 & 17.933 & *** & 0.790 & 0.624 &  &  \\
  & DCI3 & 1.264 & 0.067 & 18.804 & *** & 0.823 & 0.677 &  &  \\
  & DCI4 & 1.057 & 0.061 & 17.256 & *** & 0.765 & 0.585 &  &  \\
  & DCI5 & 1.163 & 0.063 & 18.325 & *** & 0.805 & 0.648 &  &  \\
\bottomrule
\end{tabular*}
\end{table}

\begin{table}[htbp]
\centering
\setlength{\tabcolsep}{4pt} 
\caption{Total Dataset Correlation Matrix (Diagonal: Square Root of AVE)}
\label{tab:correlation_matrix}
\begin{tabular*}{\linewidth}{@{\extracolsep{\fill}} l c c c c c c c c @{}}
\toprule
Variables                       & PD    & ID    & SD    & KD    & DT    & CSE   & CCE   & DCI   \\
\midrule
Problem-driven Design           & 0.678 &       &       &       &       &       &       &       \\
Information-driven Design       & 0.348 & 0.715 &       &       &       &       &       &       \\
Solution-driven Design          & 0.610 & 0.465 & 0.762 &       &       &       &       &       \\
Knowledge-driven Design         & 0.562 & 0.477 & 0.515 & 0.687 &       &       &       &       \\
Design thinking                 & 0.534 & 0.529 & 0.566 & 0.580 & 0.759 &       &       &       \\
Creative Self-Efficacy          & 0.459 & 0.293 & 0.383 & 0.445 & 0.429 & 0.756 &       &       \\
Collective Creative Efficacy    & 0.470 & 0.305 & 0.442 & 0.390 & 0.334 & 0.340 & 0.814 &       \\
Design Creativity \& Innovation & 0.569 & 0.375 & 0.577 & 0.509 & 0.511 & 0.569 & 0.531 & 0.791 \\
\bottomrule
\end{tabular*}
\end{table}

\begin{table}[htbp]
\centering
\setlength{\tabcolsep}{4pt} 
\caption{Group-wise Correlation Matrix (Lower Triangle: Student Group, Upper Triangle: Professional Group)}
\label{tab:correlation_matrix_groups}
\begin{tabular*}{\linewidth}{@{\extracolsep{\fill}} l c c c c c c c c @{}}
\toprule
Variables                       & PD    & ID    & SD    & KD    & DT    & CSE   & CCE   & DCI   \\
\midrule
Problem-driven Design           &       & 0.280 & 0.507 & 0.412 & 0.380 & 0.389 & 0.273 & 0.407 \\
Information-driven Design       & 0.330 &       & 0.471 & 0.537 & 0.490 & 0.185 & 0.277 & 0.301 \\
Solution-driven Design          & 0.600 & 0.368 &       & 0.369 & 0.493 & 0.286 & 0.252 & 0.487 \\
Knowledge-driven Design         & 0.581 & 0.348 & 0.486 &       & 0.418 & 0.380 & 0.210 & 0.369 \\
Design thinking                 & 0.612 & 0.529 & 0.586 & 0.655 &       & 0.362 & 0.234 & 0.377 \\
Creative Self-Efficacy          & 0.400 & 0.285 & 0.306 & 0.325 & 0.415 &       & 0.314 & 0.497 \\
Collective Creative Efficacy    & 0.473 & 0.190 & 0.384 & 0.305 & 0.321 & 0.079 &       & 0.387 \\
Design Creativity \& Innovation & 0.609 & 0.352 & 0.523 & 0.480 & 0.583 & 0.486 & 0.459 &       \\
\bottomrule
\end{tabular*}
\end{table}

\begin{table}[htbp]
\centering
\caption{Comparison of Measurement Weight Coefficients for 40 Items (Student vs. Professional Group)}
\label{tab: item_weight_comparison}
\begin{tabular*}{\linewidth}{@{\extracolsep{\fill}} ll rrrr c @{}}
\toprule
Variables & Items & {Student Group} & {Professional Group} & {Diff} & {Z-value} & Sig. \\
\midrule
\multirow{5}{*}{Problem-driven Design} 
& PD1 & 1.000 & 1.000 & 0.000 & --- & ns \\
& PD2 & 1.091 & 1.201 & -0.110 & -0.525 & ns \\
& PD3 & 1.001 & 1.283 & -0.282 & -1.309 & ns \\
& PD4 & 0.813 & 1.155 & -0.342 & -1.672 & ns \\
& PD5 & 1.093 & 1.642 & -0.549 & -2.299 & *   \\
\cmidrule{1-7}
\multirow{5}{*}{Information-driven Design} 
& ID1 & 1.000 & 1.000 & 0.000 & --- & ns \\
& ID2 & 0.660 & 0.797 & -0.137 & -1.075 & ns \\
& ID3 & 1.075 & 1.214 & -0.139 & -0.936 & ns \\
& ID4 & 0.820 & 0.937 & -0.117 & -0.812 & ns \\
& ID5 & 1.013 & 1.172 & -0.160 & -1.112 & ns \\
\cmidrule{1-7}
\multirow{5}{*}{Solution-driven Design} 
& SD1 & 1.000 & 1.000 & 0.000 & --- & ns \\
& SD2 & 1.395 & 1.272 & 0.123 & 0.685 & ns \\
& SD3 & 1.285 & 1.299 & -0.014 & -0.083 & ns \\
& SD4 & 1.335 & 1.106 & 0.229 & 1.380 & ns \\
& SD5 & 1.239 & 1.029 & 0.210 & 1.209 & ns \\
\cmidrule{1-7}
\multirow{5}{*}{Knowledge-driven Design} 
& KD1 & 1.000 & 1.000 & 0.000 & --- & ns \\
& KD2 & 0.941 & 0.843 & 0.098 & 0.670 & ns \\
& KD3 & 0.980 & 0.890 & 0.089 & 0.663 & ns \\
& KD4 & 0.714 & 0.654 & 0.060 & 0.475 & ns \\
& KD5 & 0.931 & 0.636 & 0.295 & 2.137 & *   \\
\cmidrule{1-7}
\multirow{5}{*}{Design Thinking} 
& DT1 & 1.000 & 1.000 & 0.000 & --- & ns \\
& DT2 & 1.012 & 1.006 & 0.006 & 0.055 & ns \\
& DT3 & 1.065 & 1.034 & 0.031 & 0.297 & ns \\
& DT4 & 0.551 & 0.717 & -0.166 & -1.783 & ns \\
& DT5 & 0.778 & 0.760 & 0.018 & 0.183 & ns \\
\cmidrule{1-7}
\multirow{5}{*}{Creative Self-Efficacy} 
& CSE1 & 1.000 & 1.000 & 0.000 & --- & ns \\
& CSE2 & 0.740 & 0.745 & -0.005 & -0.043 & ns \\
& CSE3 & 1.050 & 1.004 & 0.047 & 0.387 & ns \\
& CSE4 & 0.807 & 0.948 & -0.141 & -1.315 & ns \\
& CSE5 & 0.918 & 0.701 & 0.217 & 1.898 & ns \\
\cmidrule{1-7}
\multirow{5}{*}{Collective Creative Efficacy} 
& CCE1 & 1.000 & 1.000 & 0.000 & --- & ns \\
& CCE2 & 1.065 & 1.100 & -0.034 & -0.314 & ns \\
& CCE3 & 1.060 & 0.927 & 0.132 & 1.325 & ns \\
& CCE4 & 0.923 & 0.942 & -0.018 & -0.171 & ns \\
& CCE5 & 0.831 & 0.703 & 0.128 & 1.213 & ns \\
\cmidrule{1-7}
\multirow{5}{*}{\makecell[l]{Design Creativity \\ \& Innovation}} 
& DCI1 & 1.000 & 1.000 & 0.000 & --- & ns \\
& DCI2 & 1.518 & 1.090 & 0.427 & 2.442 & *   \\
& DCI3 & 1.850 & 1.197 & 0.653 & 3.289 & **  \\
& DCI4 & 1.555 & 1.034 & 0.521 & 2.916 & **  \\
& DCI5 & 1.499 & 1.185 & 0.314 & 1.763 & ns \\
\bottomrule
\end{tabular*}
\end{table}

\begin{table}[htbp]
\centering
\setlength{\tabcolsep}{6pt} 
\caption{Comparison of Structural Path Coefficients (Student vs. Professional Group)}
\label{tab: structural_path_comparison}
\begin{tabular*}{\linewidth}{@{\extracolsep{\fill}} l rrrr c @{}}
\toprule
\makecell[l]{Path \\ (Predictor $\rightarrow$ Criterion)} & {Student} & {Professional} & {Diff} & {Z-value} & Sig. \\
\midrule
\makecell[l]{Creative Self-Efficacy \\ $\rightarrow$ Collective Creative Efficacy} & -0.114 & 0.229 & -0.343 & -2.899 & ** \\ \addlinespace[2pt]
\makecell[l]{Design thinking \\ $\rightarrow$ Collective Creative Efficacy} & 0.383 & 0.153 & 0.231 & 2.128 & * \\ \addlinespace[2pt]
\makecell[l]{Knowledge-driven Design \\ $\rightarrow$ Design thinking} & 0.419 & 0.167 & 0.252 & 1.846 & ns \\ \addlinespace[2pt]
\makecell[l]{Creative Self-Efficacy \\ $\rightarrow$ Design Creativity \& Innovation} & 0.200 & 0.274 & -0.074 & -0.965 & ns \\ \addlinespace[2pt]
\makecell[l]{Solution-driven Design \\ $\rightarrow$ Design thinking} & 0.275 & 0.377 & -0.102 & -0.649 & ns \\ \addlinespace[2pt]
\makecell[l]{Design thinking \\ $\rightarrow$ Creative Self-Efficacy} & 0.357 & 0.408 & -0.051 & -0.512 & ns \\ \addlinespace[2pt]
\makecell[l]{Problem-driven Design \\ $\rightarrow$ Design thinking} & 0.292 & 0.211 & 0.081 & 0.475 & ns \\ \addlinespace[2pt]
\makecell[l]{Information-driven Design \\ $\rightarrow$ Design thinking} & 0.339 & 0.290 & 0.048 & 0.375 & ns \\ \addlinespace[2pt]
\makecell[l]{Design thinking \\ $\rightarrow$ Design Creativity \& Innovation} & 0.214 & 0.187 & 0.027 & 0.362 & ns \\ \addlinespace[2pt]
\makecell[l]{Collective Creative Efficacy \\ $\rightarrow$ Design Creativity \& Innovation} & 0.173 & 0.197 & -0.024 & -0.332 & ns \\
\bottomrule
\end{tabular*}
\end{table}

\begin{table}[htbp]
\centering
\setlength{\tabcolsep}{6pt}
\caption{Comparison of Exogenous Variances and Covariances (Student vs. Professional Group)}
\label{tab: variance_covariance_comparison}
\begin{tabular*}{\linewidth}{@{\extracolsep{\fill}} l rrrr c @{}}
\toprule
Relationship & {Student} & {Professional} & {Diff} & {Z-value} & Sig. \\
\midrule
\makecell[l]{Variance \\ (Problem-driven Design)} & 1.294 & 0.574 & 0.721 & 2.737 & ** \\ \addlinespace[2pt]
\makecell[l]{Problem-driven Design \\ $\leftrightarrow$ Solution-driven Design} & 0.688 & 0.338 & 0.349 & 2.510 & * \\ \addlinespace[2pt]
\makecell[l]{Problem-driven Design \\ $\leftrightarrow$ Knowledge-driven Design} & 0.723 & 0.364 & 0.358 & 2.291 & * \\ \addlinespace[2pt]
\makecell[l]{Information-driven Design \\ $\leftrightarrow$ Knowledge-driven Design} & 0.402 & 0.666 & -0.264 & -1.684 & ns \\ \addlinespace[2pt]
\makecell[l]{Problem-driven Design \\ $\leftrightarrow$ Information-driven Design} & 0.396 & 0.227 & 0.169 & 1.333 & ns \\ \addlinespace[2pt]
\makecell[l]{Solution-driven Design \\ $\leftrightarrow$ Knowledge-driven Design} & 0.536 & 0.380 & 0.155 & 1.114 & ns \\ \addlinespace[2pt]
\makecell[l]{Variance \\ (Solution-driven Design)} & 1.020 & 0.778 & 0.242 & 1.079 & ns \\ \addlinespace[2pt]
\makecell[l]{Variance \\ (Knowledge-driven Design)} & 1.198 & 1.343 & -0.145 & -0.484 & ns \\ \addlinespace[2pt]
\makecell[l]{Information-driven Design \\ $\leftrightarrow$ Solution-driven Design} & 0.391 & 0.443 & -0.052 & -0.408 & ns \\ \addlinespace[2pt]
\makecell[l]{Variance \\ (Information-driven Design)} & 1.113 & 1.135 & -0.022 & -0.083 & ns \\
\bottomrule
\end{tabular*}
\end{table}

\end{document}